\documentclass{biophys}
\usepackage{helvet,times}
\usepackage{bm,textcomp}

\jno{kxl014} 
\gridframe{N}
\cropmark{N}

\usepackage{amsmath}

\usepackage[round,numbers,sort&compress]{natbib}

\renewcommand{\figurename}{FIGURE}
\renewcommand{\theequation}{\arabic{equation}} 

\newcommand{\Fone}{F$_1$-ATPase}
\newcommand{\Fones}{F$_1$-ATPase }
\newcommand{\Fon}{F$_1$}
\newcommand{\Fons}{F$_1$ }
\newcommand{\Qint}{Q_{\rm int}}
\newcommand{\Qext}{Q_{\rm ext}}
\newcommand{\Swt}[2]{f_{#1} ({#2})}

\newcommand{\Swtp}[2]{R^+_{#1} ({#2})}
\newcommand{\Swtm}[2]{R^-_{#1} ({#2})}
\newcommand{\fSwtp}[2]{f^+_{#1} ({#2})}
\newcommand{\fSwtm}[2]{f^-_{#1} ({#2})}
\newcommand{\Swtpm}[2]{R^\pm_{#1} ({#2})}

\newcommand{\Pot}[2]{U_{#1} ({#2})}
\newcommand{\WPot}[2]{\widetilde{U}_{#1} ({#2})}
\newcommand{\Prob}[2]{P_{#1}^t ({#2})}
\newcommand{\Probs}[2]{P_{#1}^{*} ({#2})}
\newcommand{\Probt}[2]{P_{#1}^{\tilde{t}} ({#2})}
\newcommand{\Probz}[2]{P_{#1}^{(0)} ({#2})}
\newcommand{\Probo}[2]{P_{#1}^{(1)} ({#2})}
\newcommand{\Probss}[2]{P_{#1}^{\rm ss} ({#2})}
\newcommand{\Probeq}[2]{P_{#1}^{\rm eq} ({#2})}
\newcommand{\vvs}{v}
\newcommand{\vmax}{v_{\rm max}}
\newcommand{\Trans}[2]{\Lambda_{#1} ({#2})}

\newcommand{\mmodel}{totally asymmetric allosteric model }

\newcommand{\model}{TASAM }
\newcommand{\modelss}{TASAM}

\begin{document}

\setcounter{page}{1} 

\title{Nonequilibrium dissipation-free transport in F$_1$-ATPase and the thermodynamic role of asymmetric allosterism}

\author{Kyogo Kawaguchi,$^\ast$ Shin-ichi Sasa,$^\dagger$ and Takahiro Sagawa$^\ddagger$}

\address{$^\ast$Department of Physics, The University of Tokyo, Tokyo 113-0033, Japan; $^\dagger$Department of Physics, Kyoto University, Sakyo-ku, Kyoto 606-8502, Japan; and $^\ddagger$Department of Basic Science, The University of Tokyo, Tokyo 153-8902, Japan}



\begin{abstract}%
{\Fones (or \Fon), the highly-efficient and reversible biochemical engine, has motivated physicists as well as biologists to imagine the design principles governing machines in the fluctuating world. Recent experiments have clarified yet another interesting property of \Fon; the dissipative heat inside the motor is very small, irrespective of the velocity of rotation and energy transport. Conceptual interest is devoted to the fact that the amount of internal dissipation is not simply determined by the sequence of equilibrium pictures, but also relies on the rotational-angular dependence of nucleotide affinity, which is a truly nonequilibrium aspect. We propose that the \mmodel (\modelss), where adenosine triphosphate (ATP) binding to \Fons is assumed to have low dependence on the angle of the rotating shaft, produces results that are most consistent with the experiment. Theoretical analysis proves the crucial role of two time scales in the model, which explains the universal mechanism to produce the internal dissipation-free feature. The model reproduces the characteristic torque dependence of the rotational velocity of \Fon, and predicts that the internal dissipation upon the ATP synthesis direction rotation becomes large at the low nucleotide condition. }
{}
{kyogok@daisy.phys.s.u-tokyo.ac.jp}
\end{abstract}

\maketitle 
\section*{INTRODUCTION}
One of the major progress in modern statistical physics is the understanding of stochastic thermodynamics, where behavior of thermodynamic machines at the fluctuating world is the main target of study~\cite{sekimoto_book2010,seifert2012}.
In sharp contrast to macroscopic systems, fluctuations play significant roles in microscopic engines~\cite{reimann2002,Blickle2011}, as in the cases of biomolecular motors~\cite{howard_book2001}.
Recent studies have further clarified the role of information in thermodynamics~\cite{toyabe2010_demon,Berut2012} which provides new perspectives on the design principle of molecular motors~\cite{Mandal2012,Horowitz2012}.
In light of these developments both in theory and experiments, we are now in the position to reveal the fundamental rules that govern nonequilibrium molecular machines.

\begin{figure}[hbt]
 \begin{center}
  \includegraphics[width=87mm]{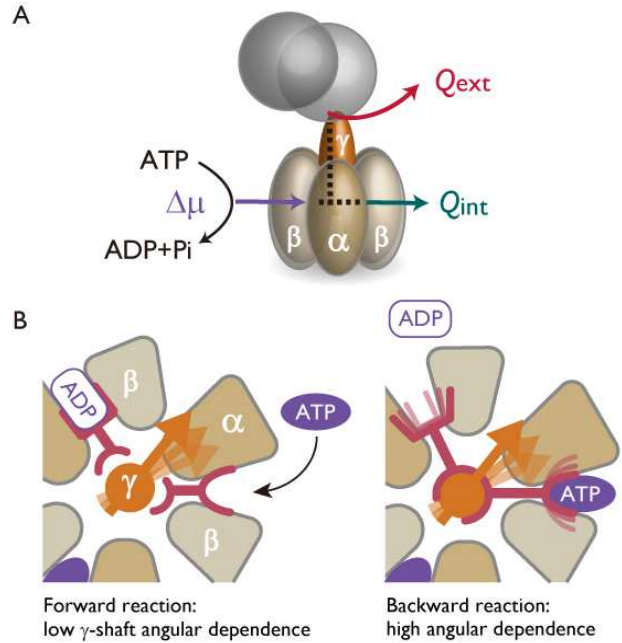}
 \end{center}
  \caption{\label{Fig1} The totally asymmetric allosteric model (TASAM) explains the internal dissipation-free feature of \Fone.
({\it A}) Summary of the energy income and outflow of \Fons in the ATP hydrolysis rotation in the absence of external load. Free energy of the chemical fuel ($\Delta \mu$) is dissipated through the rotary motion of the $\gamma$ shaft ($\Qext$), or through the $\alpha \beta$ complex where the reactions take place ($\Qint$).
({\it B}) Schematic of the \modelss. When the motor is waiting for the ATP to bind (left), the binding site is free from the control by the $\gamma$ shaft rotary angle. Conversely, once the ATP is bound to the motor and the ADP release has proceeded (right), the affinity of the binding sites to the nucleotides strictly depends on the $\gamma$ shaft angle.
}
\end{figure}

Among many molecular motors, the detailed energetics of \Fons has attracted special attention due to its important roles in the metabolic network, and its unique feature of reversibility~\cite{Boyer1997,yasuda1998,Wang1998,itoh2004}.
Rotational motion of \Fons in the ATP hydrolysis direction is roughly explained as follows: the $\alpha \beta$ subunits change their conformations following the events of the binding of ATP, the hydrolysis reaction ATP $\to$ adenosine diphosphate (ADP) + phosphate (Pi), and the release of Pi and ADP.
These chemical reaction-coupled conformation changes induce the $\gamma$ subunit (shaft) to rotate 120$^\circ$ per single ATP input and ADP+Pi output.
Energetically (Fig.~1 {\it A}), in the absence of load, the income of this motor is the chemical free energy of ATP ($\Delta \mu$), and the outflow is made either through the $\gamma$ shaft rotation ($\Qext$, external dissipation) or the switching of conformation in $\alpha \beta$ ($\Qint$, internal dissipation).

Recent precise experiments on \Fons have shown that, in the absence of external load, the chemical energy consumption is made almost 100\% through the rotational motion, $\Qext \sim \Delta \mu$, meaning that the internal dissipation is almost zero~\cite{toyabe2010_F1,toyabe2013}, $\Qint \sim 0$.
This highly suggestive result provokes us to consider the hidden rules that govern molecular motors.
In fact, the established theories in stochastic thermodynamics (e.g. the fluctuation theorem~\cite{evans1993,jarzynski1997,kurchan1998} and the second law) prove that the total energy dissipation is given by $\Delta \mu  = \Qext + \Qint$, but cannot explain any restrictions which rule the individual values of $\Qext$ and $\Qint$.
The experiment shows the further surprising aspect of \Fon; the internal dissipation-free feature does not depend on the rotational velocity~\cite{toyabe2010_F1}.
Conceptual interest lies in the fact that these observed data reflect the truly nonequilibrium character of \Fon, and include more information than the equilibrium; the internal dissipation is a quantity which cannot be determined even if the full free energy landscape is measured or calculated.

Interestingly, the chemical reactions, which are the main source of the rotational motion, are itself controlled by the rotary position of the $\gamma$ shaft \cite{watanabe2012,adachi2012}, which is a feature known as allosterism.
The amount of internal/external dissipation in a model may be manipulated by setting various types of $\gamma$ angular dependence of the chemical reactions.
In \cite{zimmermann2012}, a load-sharing factor was introduced as a parameter for this angular dependence. It was observed that in the case where the load-sharing is assumed to be largely asymmetric, numerical simulation produces consistent results with the internal dissipation-free feature of \Fone~\cite{toyabe2010_F1}.

Following the approach taken in \cite{zimmermann2012}, we reconsider the features of \Fons at the level of the phenomenological model, where only the rotary degree of freedom of the $\gamma$ shaft and the nucletide state of the $\alpha \beta$ subunits are taken into account.
Specifically, we introduce the \mmodel (\modelss, Fig.~1 {\it B}) in order to explain the internal dissipation-free nature of \Fon.
The key assumption in this model is that the ATP binding to the motor, which is the limiting slow process, is allowed with equal probability over the $\gamma$ shaft angle.
The motor thus passively waits and lets the ATP to bind freely, and decides whether or not to release this ATP depending on the angle of the $\gamma$ shaft.
As we shall see in detail (Fig.~2 {\it C}), numerical results in our model show that $\Qext$ is kept close to $\Delta \mu$ for a broad range of rotational velocity.
We identify the crucial role of two time scales in the TASAM, which explains the universal mechanism in the model to produce the internal dissipation-free feature.
Moreover, we analyze the nucleotide concentration dependence on the torque-velocity relation (Fig.~5), and find that a certain asymmetric pattern observed in experiment is reproduced by the TASAM. 

Our stochastic model is based on confirmed properties of \Fon, such as the discrete steps~\cite{masaike2008,yasuda2001}, mechanical potentials~\cite{toyabe2012}, and large stall force~\cite{toyabe2011stall} with all the real parameters.
However, we do not refer to the microscopic interactions, for example, at the amino acid residue level, which is typically required in molecular dynamic simulations.
Nevertheless, the key feature of the \Fons energetics and dynamics seem to be well reproduced by the simple one-dimensional description.
The existence of such consistent description encourages us to consider the fundamental design principle behind molecular machines, since at this coarse-grained scale, comparison between different bio-motors (including linear processive motors) and blueprints of future nanomachines is possible.

\section*{MATERIALS AND METHODS}
\section*{Model description}
We introduce a simple one-dimensional model based on the Brownian motion and potential switching scheme. The mechanical potentials $\Pot{n}{x}$ trap the rotational degree of freedom $x$ of the probe attached to the tip of the $\gamma$ subunit. Here, $n=0,\pm 1, \pm 2,...$ are the number of ATP consumed through the rotary motion. Each potential is created by the interaction between the $\gamma$ subunit and the $\alpha \beta$ subunits~\cite{masaike2008,toyabe2012}, and the joint between the $\gamma$ subunit and the probe bead~\cite{okuno2010}.

Due to the three-fold symmetry of the motor (Fig.~1), the potentials are translationally identical $\Pot{n}{x}=\Pot{0}{x-120^\circ \times n}$.
Since the substeps corresponding to the hydrolysis/synthesis reactions + releasing/binding of Pi are fast~\cite{yasuda2001}, the hydrolysis dwell potentials are effectively included in $\Pot{n}{x}$, and the ATP binding followed by ADP release is considered as the rate-limiting step in the full reaction scheme. By assuming that the ATP binding dwell and the hydrolysis dwell potentials are harmonic  with the same spring constants~\cite{watanabe2012}, $k$, the form of $\Pot{n}{x}$ is estimated as an effective potential, $U_0(x) /k_{\rm B} T=kx^2/2 - \log[\exp(-klx) + \exp(\widetilde{\Delta} \mu/k_{\rm B}T + kl^2/2) ]$. Here, $l=40^\circ$ is the angle of the substep, and $\widetilde{\Delta} \mu$ is the free energy difference between the ATP hydolysis dwell state and the binding dwell state. 
By fitting the data in \cite{toyabe2012}, we obtained $k=0.0061$ deg$^{-2}$ and $\widetilde{\Delta} \mu = 5.2 k_B T$, which are in good agreement with previous independent observations~\cite{watanabe2012,okuno2010,yasuda2001}. See Supporting Material A for the derivation of $U_0(x)$.

The angular position of the probe bead, $x$, undergoes an overdamped Brownian motion inside each potential:
 \begin{eqnarray}
 \Gamma \dot{x}  = -\frac{\partial }{\partial x} U_n(x) -F + \sqrt{2\Gamma k_{\rm B} T} \xi_t, \label{LE}
 \end{eqnarray}
where $\xi_t$ is the Gaussian white noise with unit variance, $k_{\rm B}$ is the Boltzmann constant, $F$ is the applied external torque, $\Gamma$ is the friction coefficient determined by the size of the probe bead, and $T$ is the temperature of the water (i.e., the heat bath).
The potentials are switched according to chemical-reaction induced Poissonian transitions (Fig.~2 {\it A}).
The switching rate in the forward direction ($n \to n+1$), $\Swtp{n}{x}$, corresponds to the ATP binding + ADP release reaction, and the rate in the backward ($n \to n-1$) direction, $\Swtm{n}{x}$, corresponds to the ADP binding + ATP release, where we assume $\Swtpm{n}{x}=\Swtpm{0}{x-120^\circ \times n}$.
The switching rate functions satisfy the local detailed balance condition (Fig.~2 {\it A})
 \begin{eqnarray}
 \frac{\Swtp{n}{x}}{\Swtm{n+1}{x}} = \exp \left\{ \frac{1}{k_{\rm B} T}\left [U_n(x) - U_{n+1} (x) + \Delta \mu \right] \right\}, \label{DB}
 \end{eqnarray}
where we used the free energy difference between a single molecule ATP and ADP+Pi, $\Delta \mu = \Delta \mu_0 + k_{\rm B} T \log ([{\rm ATP}][{\rm H_2 O}]/[{\rm ADP}][{\rm Pi}])$.
The finite free energy input $\Delta \mu$ in Eq.~{\ref{DB} lets the motion of $x$ to show a finite steady-state current in the plus direction, whereas at $\Delta \mu =0$ the rotational motion stalls and the dynamics is in equilibrium.
The conditon Eq.~\ref{DB} also guarantees that the rotational motion would stall when the external torque $F=\Delta \mu /L$ is applied to the motor, and would rotate backwards (i.e., the ATP synthesis rotation) when $F>\Delta \mu /L$, corresponding to the tight-coupling feature of \Fons confirmed in the experiment~\cite{toyabe2011stall}.

Here we note on the requirement to introduce our potential switching model. Although it has been clarified that the discrete position of the gamma subunit ($0^\circ, 80^\circ, 120^\circ...$) is coupled strongly to the nucleotide state and conformation of the alpha-beta subunits \cite{yasuda1998,masaike2008,yasuda2001}, what is happening in between these discrete angles is a question with yet no concrete answer. In fact, it has been clarified in \cite{watanabe2012,adachi2012} that the chemical reactions do not necessarily occur precisely at a certain angle of the $\gamma$ shaft, meaning that the timing of chemical reaction is indeed a stochastic phenomena with respect to the the angular position. Furthermore, it was shown in Ref.~\cite{toyabe2012} that there exist separate potentials corresponding to the discrete positions with a certain extent of overlap. Taking these facts together, it is required to adopt our model depicted in Fig.~2A to reproduce the experimentally observed kinetics and energetics of the \Fons motor.

In the stochastic dynamics of our model, there are two paths for the motor to exchange energy with the surrounding water (Fig.~2{\it A}).
One is through the change of rotational position, which corresponds to the external dissipation, since this energy flows out from the $\gamma$ shaft.
The other is through the change of mechanical potential, which is the internal dissipation corresponding to the energy used to change the conformation of $\alpha\beta$.
To characterize these quantities, we introduce the probability density function $\Probss{n}{x}$, which describes the steady-state probability of $x$ under the condition that the trapping potential is $U_n(x)$.
Note that $\Probss{n}{x}=\Probss{0}{x-120^\circ \times n}$ due to the translational symmetry.
We further define $\Trans{n}{x}:= \Probss{n}{x} \Swtp{n}{x} -  \Probss{n+1}{x} \Swtm{n+1}{x}$, which is the steady-state switching rate that characterizes the switching position. The first term in the right-hand side corresponds to the probability density of the position at which the forward switching ($n\to n+1$) occurs, and the second term corresponds to that of the backward switching ($n+1 \to n$). By integrating $\Trans{n}{x}$, we obtain the net transport rate from potential $n$ to $n+1$,  $\vvs:=  \int ^\infty _{-\infty} dx \Trans{n}{x}/3$, which is the steady-state rotational velocity.

The heat dissipations $\Qint$ and $\Qext$ in the model are defined as the steady-state average of the internal and external heat dissipation per 120$^\circ$ step, respectively (Fig.~2 {\it A}).
$\Qint$, the energy dissipation accompanying the switching of $n$, is calculated by
\begin{eqnarray}
 \Qint:= \frac{1}{3 \vvs} \int dx \Trans{n}{x} \left[ \Pot{n}{x}-\Pot{n+1}{x} + \Delta \mu \right],\label{qint}
\end{eqnarray}
where $ \Pot{n}{x}-\Pot{n+1}{x} + \Delta \mu $ corresponds to the total energy shift upon the potential switching $n \to n+1$ at position $x$, and the average is taken with respect to the  probability distribution of the switching position.
$\Qext$, the dissipation through the spatial motion of $x$, is the difference between the total dissipation and the internal one:
\begin{eqnarray}
 \Qext:=  |\Delta \mu - FL| -\Qint . \label{qext}
\end{eqnarray}
Here, $FL$ is the work performed by the motor against the external torque per forward step, and therefore $|\Delta \mu - FL|$ corresponds to the total dissipation per step. We take the absolute value since the average  stepping direction changes its sign at $FL=\Delta \mu$.

Since direct measurement of $\Qext$ is technically challenging, it has been estimated through indirect measurement~\cite{toyabe2010_F1,toyabe2013}. The key idea in these experiments is that $\Qext$ is related to the extent of the violation of the fluctuation response relation~\cite{haradasasa2005}, and thus may be quantified through the measurement of the angular velocity fluctuation and the response to small perturbative torque. Using this method, $\Qext$ has been obtained without referring to $\Delta \mu$ or the detail of the mechanical potential~\cite{toyabe2010_F1,toyabe2013}.

Data of rotational velocity and external heat dissipation (Figs.~2 and 5) and the switching position probability densities (Fig.~3) were obtained by numerically solving the Fokker-Planck equation corresponding to the model [Eqs.~\ref{LE} and \ref{qmodel}]. We used the diffusion constant $k_{\rm B} T / \gamma L^2 = 3.3$ sec$^{-1}$ which was noted in \cite{toyabe2012}. $\Delta \mu$ was set as 18.3 $k_{\rm B} T$ in Figs.~2, 5, and 16.5 $k_{\rm B} T$ in Fig.~3, corresponding to the experiments \cite{toyabe2010_F1}, \cite{toyabe2013}, and \cite{toyabe2012}, respectively.

\begin{figure*}[!h]
 \begin{center}
  \includegraphics[width=178mm]{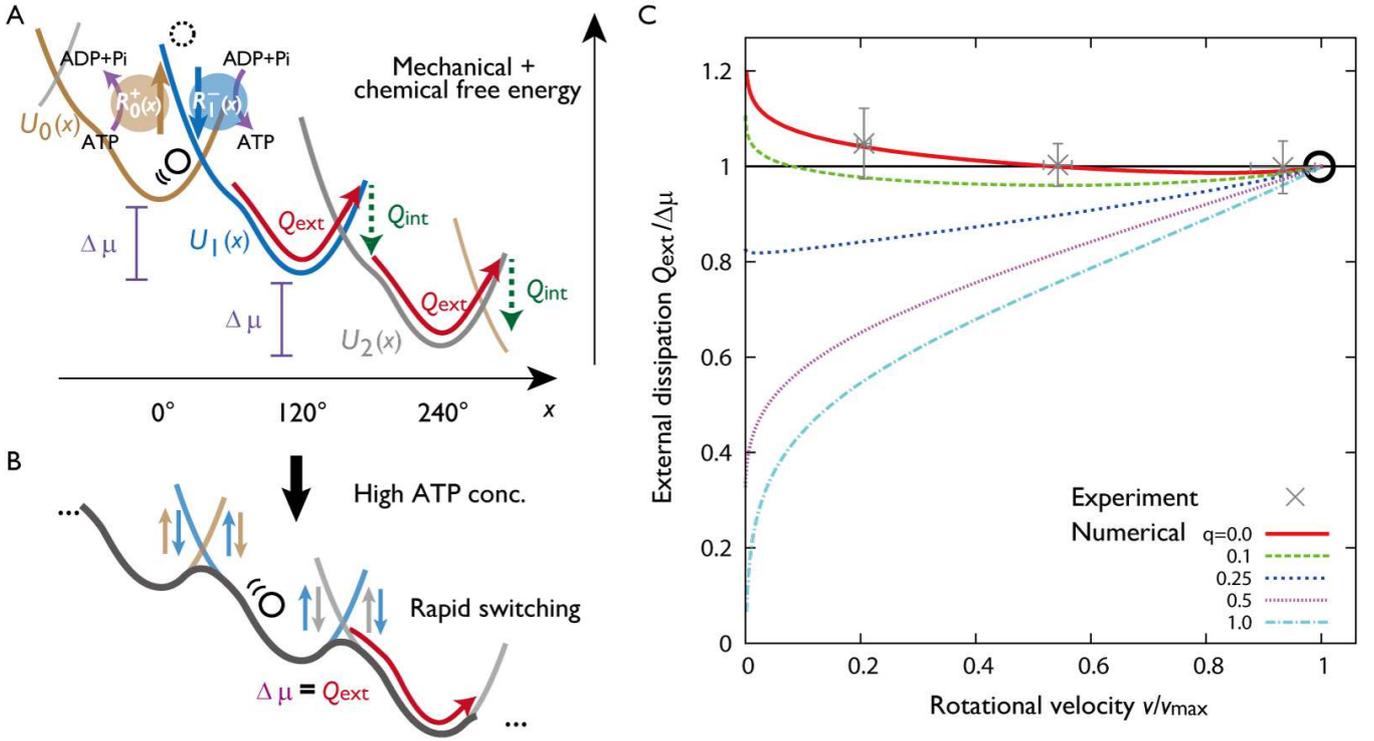}
 \end{center}
  \caption{\label{Fig2} External heat dissipation defined in the molecular motor model and its dependence on transport velocity.
({\it A}) Potential switching model and schematic of the heat dissipation for the hydrolysis-driven rotation of \Fons in the absence of applied external torque ($F=0$). The black circle represents the angular position of the probe bead attached to the $\gamma$ subunit. The kinetics of $x$ is described by an overdamped Brownian motion inside each potential. Potentials are switched according to the angular position dependent rates $\Swtpm{n}{x}$.
({\it B}) Effective potential in the high ATP concentration limit.
Since the switching dynamics is fast, the independent potentials become invisible, and the dynamics follows the tilted periodic potential description, irrespective of the form of $\Swtpm{n}{x}$.
In this limit, the external dissipation $\Qext$ becomes equal to $\Delta \mu$.
({\it C}) Rotational velocity $v$ versus the external heat dissipation per step $\Qext$.
The red line is the case of the \model ($q=0$), and the other lines correspond to various $q\neq 0$ models introduced by Eq.~\ref{qmodel}.
Experimental data were obtained from Ref.~\cite{toyabe2010_F1} (errorbars: standard error of mean). The black circle indicates the high ATP concentration limit.
}
\end{figure*}

\section*{RESULTS}

First, we show in the model that the steady-state rotation rate $\vvs$ converges to the finite maximum velocity $v_{\rm max}$, and the external dissipation $Q_{\rm ext}$ converges to $\Delta \mu$, in the high $[{\rm ATP}]$ (with fixed $\Delta \mu$) situation.
To see this, let us write $R^{\pm}_n (x) = W f ^{\pm}_n (x)$, where $W$ is the rate characterizing
the chemical reaction, and $f^{\pm}_n(x)$ are dimension-less functions which do not explicitly depend on the nucleotide and phosphate concentrations.
Here, $W$ is an increasing function of the ATP concentration, which should be roughly expressed as $W \simeq k_{\rm ATP}$ [ATP], although we do not assume a certain value for $k_{\rm ATP}$.
When $W$ is sufficiently large (see Supporting Material), we can prove that the dynamics becomes independent of the form of $f ^{\pm}_n (x)$, and the motion of the probe is then described by the one-dimensional Brownian motion in an effective tilted periodic potential (Fig.~2 {\it B}).
This means that the switching of mechanical potentials becomes too fast to be observed as distinct steps.
$v_{\rm max}$ corresponds to the steady state velocity determined by this effective potential.
Velocity saturation at high ATP concentration is a well-known property in molecular motors, which has been phenomenologically understood through the Michaelis-Menten curve~\cite{howard_book2001}.
Since in this limit the potential switching dynamics is sufficiently fast and reaches equilibrium, the energy dissipation accounting for the $\alpha \beta$ conformation change is balanced and becomes zero, $\Qint =0$.
Therefore, the dissipation through the rotational motion equals the free energy input $Q_{\rm ext}  =  \Delta \mu$, in the absence of torque.
This result is also consistent with the observation in the experiment~\cite{toyabe2013}, where the external force dependence of $Q_{\rm ext}$ was measured under the condition of high nucleotide concentrations (see also Fig.~5 {\it B}).

Now, the torque-free experiment~\cite{toyabe2010_F1} shows that even when [ATP] is low enough and $v < v_{\rm max}$, the external dissipation $\Qext$ is still close to $\Delta \mu$ (Fig.~2 {\it C}, crosses).
Although we have shown that the feature where $\Qext \sim \Delta \mu$ is achieved for any switching rate function when $v\sim v_{\rm max}$, the external dissipation has a strong dependence on the functional form of $f_{n}^{\pm}(x)$ in the low velocity regime, since the value of $\Qint$ (and consequently, $\Qext$) is determined by the typical position $x$ at which the switching occurs (Fig.~2 {\it A}).
The experimental observation is striking, since the individual functional forms of $f_{n}^{\pm}(x)$ are arbitrary as long as the detailed balance condition Eq.~\ref{DB} is satisfied; there is no general reason for $\Qext$ to become close to $\Delta \mu$ in the low velocity regime.

\begin{figure}[!tb]
 \begin{center}
  \includegraphics[width=87mm]{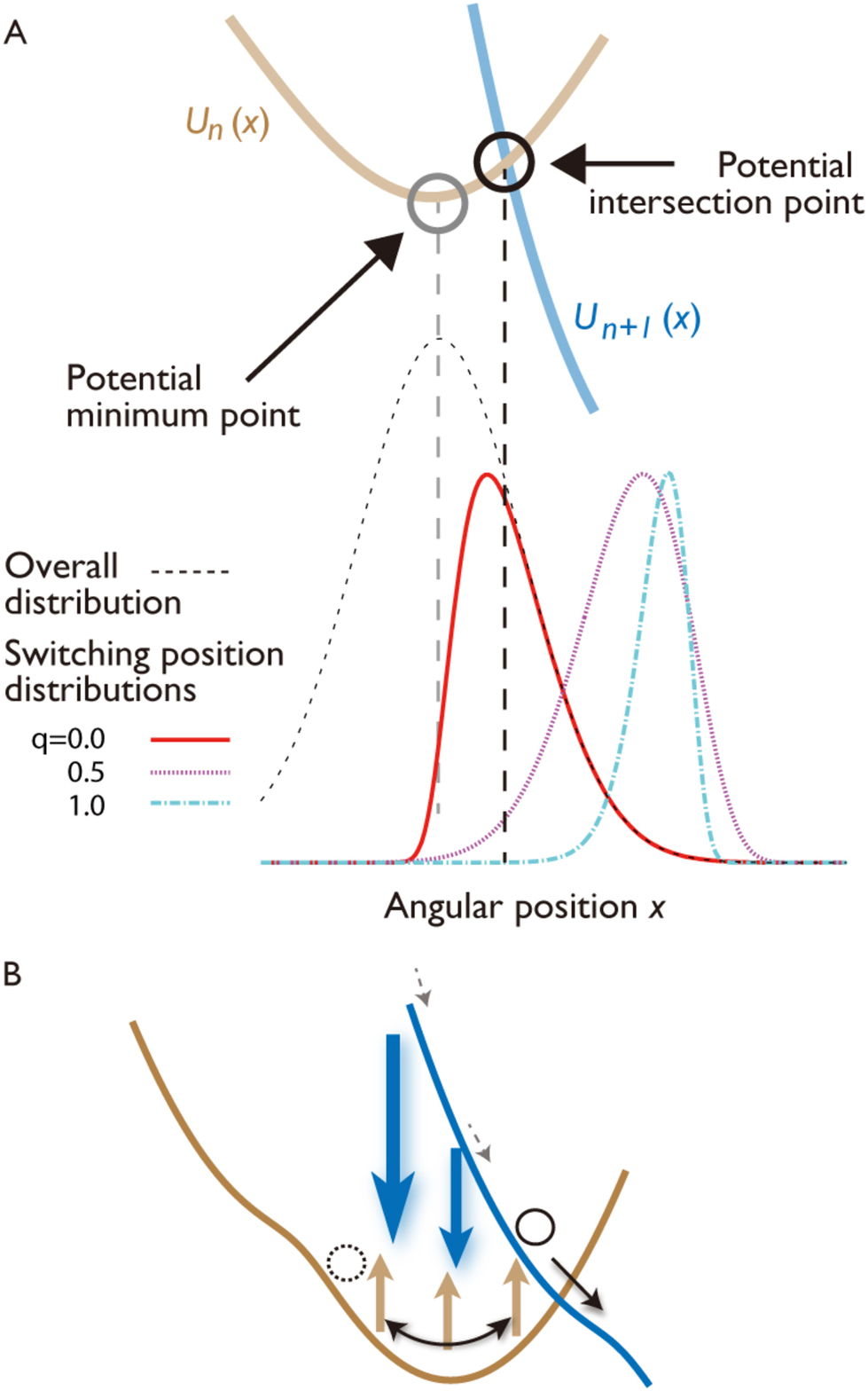}
 \end{center}
  \caption{\label{Fig3} Switching position distribution of the TASAM.
({\it A}) Distribution of the switching angular position at the slow rotation rate condition. Numerically obtained switching position distribution in the $q=0$ (red), 0.5 (pink), and 1 (light blue) models, compared for the same rotational rate $v=2$ Hz (low velocity). The vertical axis is arbitrarily scaled. Although the steady-state distribution of the $q=0$ model (black dotted line) has a peak at the minimum point of $U_n(x)$, the switching position density clearly has a peak around the intersection point, which is consistent with experiment~\cite{toyabe2012}.
({\it B}) Mechanism behind the internal dissipation-free feature of the TASAM. The potential switching may occur at any angle in the TASAM (brown arrows), while the backward switching of the potential follows instantaneously if the energy required for the forward switching was too large (blue arrows). Suppression of switching at high energy difference positions lets the switching to occur only around the potential intersection point, leading to the $\Qint \sim 0$ feature.}
\end{figure}

To see the significance of the experimental result for lower velocities, and demonstrate the uniqueness and validity of the \model (Fig.~1 {\it B}), we parameterize the switching rate functions by introducing a parameter $q$ ($0\leq q \leq 1$):
\begin{eqnarray}
\begin{split}
f_{n}^{+}(x) &=& \exp \left\{ \frac{q}{k_{\rm B} T}\left[ U_n(x) - U_{n+1} (x)+ \Delta \mu \right ]   \right\}, \\
f_{n+1}^{-} (x) &=& \exp \left\{ \frac{q-1}{k_{\rm B} T}\left[ U_n(x) - U_{n+1} (x) + \Delta \mu \right ] \right\}. 
\end{split}
\label{qmodel}
\end{eqnarray}
This $q$ determines the asymmetry in the $x$-dependence of the forward and backward switching rates, while respecting the detailed balance condition.
The \model is the case where the switching rate in the forward direction (ATP binding rate) has no dependence on the $\gamma$ angle, and therefore, corresponds to $q=0$.
The case of $q=1$ corresponds to another type of totally asymmetric model that is opposite to the TASAM, where the ATP binding event is strictly controlled by the $\gamma$ angle, whereas the reverse reaction (ADP binding) is instead independent of the angle.
In this manner, $q$ describes the asymmetry in the extent of coordination between the angle of $\gamma $ subunit and the ATP binding site.

Fig.~2 {\it C} shows the numerically obtained relation between $v$ and $\Qext$ for various $q$'s.
The \model produces the internal dissipation-free feature ($\Qext\sim \Delta \mu$), which indeed has low dependence on the rotational velocity in the broad range tested in the experiment (Fig.~2 {\it C}, red).
This feature of the model is preserved for the larger $\Delta \mu$ case (Supplemantary Material Fig.~S9).
Note that under a fixed functional form of $U_n(x)$, the maximum velocity $\vmax$ does not depend on $q$, which is why we may compare $\Qext$ between various models for the same $v/\vmax$.
Although all models produce $\Qext=\Delta \mu$ for $v=v_{\rm max}$ (black circle), the values of $\Qext$ for the cases of $q>0.1$ deviate from $\Delta \mu$ significantly at lower $v$.
Thus, we propose that the internal dissipation-free motor is obtained only by assuming low coordination  ($q\sim 0$)  in the ATP free state and high coordination in the ATP bound state.

As seen in Eq.~\ref{qint}, the value of $\Qint$ is determined by the typical switching position.
We compare $\Trans{n}{x}$ for the various models in the low velocity regime in Fig.~3 {\it A}.
Although the steady-state distribution of the $\gamma$ angle [$\Probss{n}{x}$] has a peak at the potential minimum point in this low velocity regime (Fig.~3 {\it B}, black dotted line), the peak of $\Trans{n}{x}$ in the $q=0$ model is positioned close to the point $x=x_{c,n}$ where $\Pot{n+1}{x_{c,n}}-\Pot{n}{x_{c,n}}-\Delta \mu=0$, which we shall refer to as the potential intersection point.
The mechanism which produces this behavior is that, although the ATPs most likely approach the motor when the $\gamma$ angle is around the potential minimum point, the bound ATP is almost always kicked out instantaneously, due to the large energy difference required to switch the potential at such points, $\Pot{n+1}{x}-\Pot{n}{x}-\Delta \mu \gg k_{\rm B} T$ (Fig.~3 {\it B}).
The switching would inevitably occur at positions close to $x_{c,n}$,  resulting in low energy dissipation in the potential switching (low $\Qint$).
$\Trans{n}{x}$ of the \model (Fig.~3 {\it A}) agrees with the estimated distribution presented in \cite{toyabe2012} whereas the $q=0.5$ and 1 cases (Fig.~S6) have peaks clearly shifted more toward the forward direction than the intersection point.

\begin{figure}[b]
 \begin{center}
  \includegraphics[width=87mm]{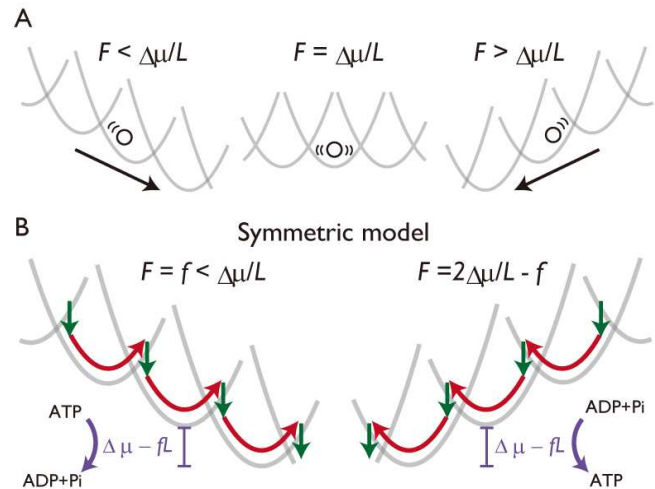}
 \end{center}
  \caption{\label{Fig4} 
Property of the potential switching model under application of external force.
({\it A}) Schematic of the torque $F$ dependence of the potential switching model. When the mechanical potential is harmonic, the model becomes symmetrical about the change $(F,q,x) \to (2\Delta \mu/L -F, 1-q, -x)$.
({\it B}) In the case of $q=0.5$, the absolute velocity and the balance between the internal/external dissipations become equivalent in the two cases, $F=f$ and $2\Delta \mu/L - f$.}
\end{figure}

On theoretical grounds, the low velocity dependence of $\Qext$ could be understood through the existence of the time scale $\tau_{\rm p}$, which determines the minimal value of $W$ at which the model presents the internal dissipation-free feature.
We have numerically verified that in the $q=0$ model, $\tau_{\rm p}$ is the time scale of the angular position to relax inside a single potential (see Supporting Material).
The crucial point is that this time scale $\tau_{\rm p}$ is much larger than the time scale $\tau_{\rm v}$, which determines the rotational velocity saturation and the single effective potential description (Fig.~2 {\it B}).
The large separation between the values of $\tau_{\rm p}$ and $\tau_{\rm v}$ causes the experimentally accessible region of the value of $W$ (corresponding to the ATP concentration) to fit inside the inequality $\tau_{\rm p}^{-1} < W \leq \tau_{\rm v}^{-1}$, and the internal dissipation-free feature is observed at all velocity conditions as a consequence.
In more practical words, our theory shows that when the actual motor adopts the \model, $\Qext\sim \Delta \mu$ holds if [ATP] is as large as 0.1 $\mu$M for the 0.3 $\mu$m probe bead case~\cite{toyabe2010_F1,toyabe2013}.
Since the velocity saturation occurs at [ATP] $>$ 5 $\mu$M under this consdition~\cite{yasuda2001}, this lower bound concentration to observe $\Qext \sim \Delta \mu$ is significantly small.

We next show how other features of the TASAM appear in measurable quantities.
For the sake of schematic explanation, let us consider that the mechanical potentials are harmonic, $U_0(x)= K(x/L)^2/2$ (Fig.~4 {\it A}). Then, the potential switching model would be symmetric about the change of parameters and coordinate, $(F,q,x ) \to (2\Delta \mu /L -F , 1-q, -x )$.
It follows from this symmetric property that in the case of $q=0.5$ (symmetric model, Fig.~4 {\it B}), the torque dependence of the steady-state rotational velocity, $v(F)$, would show an anti-symmetric curve with respect to the $F=\Delta \mu /L$ line.
Indeed, this property is observed even when we assume the non-harmonic and realistic form of $U_0(x)$ (see Fig.~S7 in Supporting Material).
On the contrary, in the TASAM ($q=0$), this feature is lost especially at low $W$, and the torque dependence of the velocity becomes sharper at $F>\Delta \mu /L$ than at $F<\Delta \mu /L$ [$v(0)= 0.32, 1.9, 7.1$ Hz lines in Fig.~5 {\it A}].
Remarkably, this feature of the TASAM matches with the experimental observation \cite{toyabe2011stall} (line points in Fig.~5 {\it A}), whereas other models fail to capture the observed feature of \Fons (Fig.~S7).

We predict that similar difference between the torque-free and torque-applied cases would be observed in the internal dissipation, if the TASAM is adopted in the \Fons motor.
In fact, Fig.~5 {\it B} shows that the internal dissipation-free nature of the TASAM is lost, especially when large torque $F>\Delta \mu/L$ is applied in the model.
The low torque dependence of $\Qint$ at high nucleotide concentration has recently been measured and reported \cite{toyabe2013}, which is consistent with the case of $v(0) = 12$ Hz in Fig.~5 {\it B}.
The validity of our model can be checked by further measuring $\Qint$ in the low nucleotide concentration condition [$v(0)= 0.32, 1.9$ and 7.1 Hz cases in Fig.~5 {\it B}].

The character of the torque dependence of TASAM could be understood through the property of the potential switching model discussed above. The dynamics of the TASAM at the presence of large applied torque, for instance $F = 2\Delta \mu/L$, becomes equivalent to that of the $q=1$ model with $F=0$ with opposite velocity.
As we have seen in Fig.~2 {\it C} and 3 {\it A}, the internal dissipation is large in the $q=1$ case, which explains why we observe large internal dissipation in the presence of torque in the TASAM at low $W$.
In this sense, adopting the TASAM in the forward step mechanism (ATP hydrolysis) is equivalent to adopting the $q=1$ model in the backward step (ATP synthesis, Fig.~5 {\it C}).
The absolute velocity at low nucleotide condition becomes faster at $F= 2 \Delta \mu/L$ than at $F=0$ in the TASAM (Fig.~5 {\it A}), which is suggestive since the \Fons is forced to rotate in the ATP hydrolysis direction in biological conditions. 
It shall be interesting to quantify how asymmetric the thermodynamic quantities and force-velocity relations are in other molecular motors.

Another way to verify our model is to directly estimate the switching rate functions.
Recently, two groups have reported the angular position dependence of chemical reaction rates in the \Fons motor~\cite{Iko2009,watanabe2012,adachi2012}. By assuming that the ATP-on rates measured in these reports correspond to the forward switching rate in our model [$R^{+}_n(x)$], we obtain $q= 0.07 \sim 0.12$ when the harmonic potential model $U_0(x) = K(x/L)^2/2$ with $K=50 k_{\rm B} T$ is adopted (see Supplementary Material B). This means that the angular position dependence of the ATP binding observed in experiment is relatively low compared with that of the ATP synthetic reaction [$R^{-}_n(x)$, corresponding to the combination of ADP-on and ATP-off], which is consistent with our assumption in the TASAM. We predict that if the direct measurement of $R^{-}_n(x)$ is possible, one should find a large dependence on the angle, since $0 \sim q \ll 1-q \sim 1$.

\begin{figure}[!h]
 \begin{center}
  \includegraphics[width=87mm]{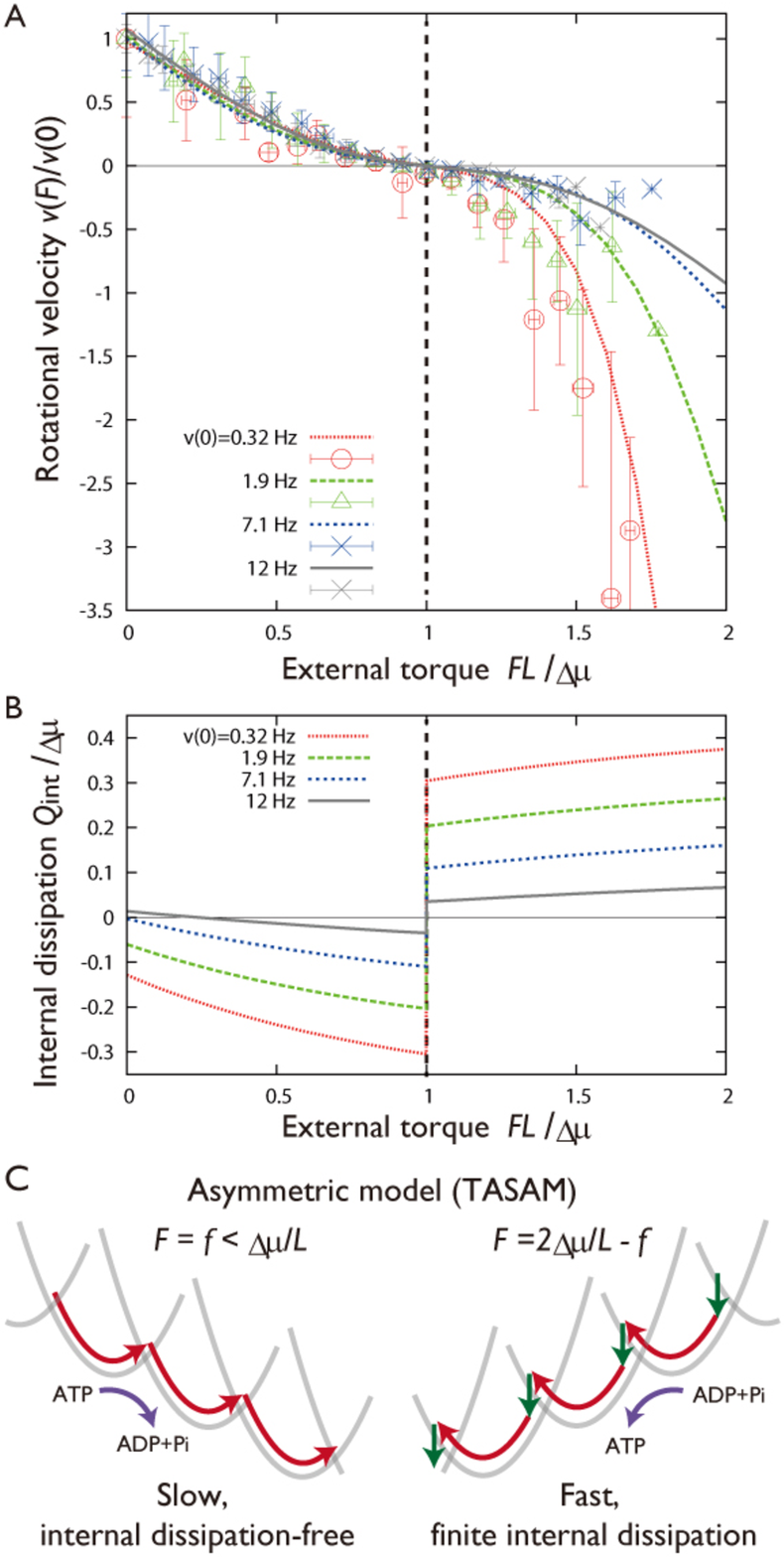}
 \end{center}
  \caption{\label{Fig5} 
Characteristic torque dependence of velocity and dissipation in the TASAM.
({\it A}) External torque dependence of the rotational velocity in the TASAM (numerical, solid lines) and the corresponding values obtained from experiment (lines and points, \cite{toyabe2011stall}). Velocity is normalized using its value at $F=0$, which is shown as the title of each lines. The $v= 0.32, 1.9, 7.1$, and 12 Hz cases correspond to $W=1.7, 15, 120$ and 1000 sec$^{-1}$ in the model, respectively, with other parameters fixed.
({\it B}) External torque dependence of the internal dissipation $\Qint$ in the TASAM. At low nucleotide conditions (small $W$), $\Qint$ significantly deviates from zero at the presence of large torque.
({\it C}) ATP hydrolysis and synthetic rotation cases in the TASAM at the low nucleotide concentration condition. In contrast to the symmetric model (Fig.~4 {\it B}), the absolute velocity and the value of internal dissipation are different in the $F=f$ and $2\Delta \mu/L -f $ cases.}
\end{figure}

\section*{DISCUSSIONS}

Let us remark on the possibility of another model to explain the experimental results.
Instead of the form we introduced in Eq.~\ref{qmodel}, we may choose a switching rate function which has a sharp peak at the position of the potential intersection point (see Supporting Material).
In such model, the peak position of $\Trans{n}{x}$, which is confined around the intersection point, would be independent of $W$, and the velocity independent feature of $\Qint \sim 0$ would be trivially obtained. The first problem of this model is that it is difficult to explain the $\Delta \mu$ independence of small internal dissipation observed in \cite{toyabe2010_F1}. Since the intersection point is $\Delta \mu$-dependent, $\Swtp{n}{x}$ must be explicitly changed depending on $\Delta \mu$ to preserve the intersection point switching.
The second problem for this model is that the $v(F)$ curve would be anti-symmetric for general $W$, opposed to the experimental observation (Fig.~S8). This is because when $R^+_n(x)$ has a sharp peak at the intersection point, $R^-_{n+1}(x)$ would also have a sharp peak around the intersection point, and the model would effectively be symmetric about the parameter and coordinate change $(F,x) \to (2\Delta \mu/L -F, -x)$.

The motor conformation dependence of the chemical reaction has been understood as the diffusion-catch mechanism in various molecular motors~\cite{toyabe2012,Iwaki2009,DeWitt2012}.
According to our theory, one way to realize such diffusion-catch feature is to simply make the chemical reaction that sets forward the desired motion to have flat dependence on the motor conformation.
The potential switching scheme is a general model that appears for example in the context of electron transfer reactions~\cite{Marcus1993}. Therefore, it is of interest to find the applicability of the two scenarios of intersection point switching: the fast switching limit and the TASAM, in a wide range of problems in biological and physical chemistry.

In summary, we showed that the general model for molecular motors is internal dissipation-free, but only when the chemical fuel concentration is high enough for the motor to reach maximum velocity.
The \model was introduced in order to explain the internal dissipation-less feature of \Fons at the small velocity condition.
We showed the consistency of the model with the experimentally observed feature in rotational velocity and the angular dependent chemical reactions, and further discussed on the possibility of large internal dissipation in the ATP synthetic rotation to be observed in future experiments.
Since the basic scheme we considered is generic, and the assumption made on the \model is simple, we consider that our theory would serve as a prototype for physicists as well as biologists to further dig into the thermodynamics of nano-machines

\section*{ACKNOWLEDGMENTS}
We thank S. Ito, Y. Nakayama, M. Sano, S. Toyabe, and E. Muneyuki for fruitful discussions. The experimental values plotted in Fig.~4 {\it A} were kindly provided by S. Toyabe. This work was supported by the JSPS Research Fellowships for Young Scientists, JSPS KAKENHI Grant Nos. 25800217, 22340114, 2234019, and 25103002, the JSPS Core-to-Core program ``Nonequilibrium dynamics of soft matter and information'', and by the Platform for Dynamic Approaches to Living System from MEXT, Japan.
\bibliography{2013kawaguchi}

\begin{thebibliography}{31}
\providecommand{\url}[1]{\texttt{#1}}
\providecommand{\urlprefix}{ }

\bibitem[Sekimoto(2010)]{sekimoto_book2010}
Sekimoto, K., 2010.
\newblock Stochastic Energetics (Lecture Notes in Physics).
\newblock Springer, Berlin.

\bibitem[Seifert(2012)]{seifert2012}
Seifert, U., 2012.
\newblock Stochastic thermodynamics, fluctuation theorems and molecular
  machines.
\newblock \emph{Rep. Prog. Phys.} 75:126001--126058.

\bibitem[Reimann(2002)]{reimann2002}
Reimann, P., 2002.
\newblock {Brownian motors : noisy transport far from equilibrium}.
\newblock \emph{Phys. Rep.} 361:57--265.

\bibitem[Blickle and Bechinger(2011)]{Blickle2011}
Blickle, V., and C.~Bechinger, 2011.
\newblock {Realization of a micrometre-sized stochastic heat engine}.
\newblock \emph{Nature Physics} 8:143--146.

\bibitem[Howard(2001)]{howard_book2001}
Howard, J., 2001.
\newblock Mechanics of Motor Proteins and the Cytoskeleton.
\newblock Sinauer, Sunderland.

\bibitem[Toyabe et~al.(2010{\natexlab{a}})Toyabe, Sagawa, Ueda, Muneyuki, and
  Sano]{toyabe2010_demon}
Toyabe, S., T.~Sagawa, M.~Ueda, E.~Muneyuki, and M.~Sano, 2010.
\newblock Experimental demonstration of information-to-energy conversion and
  validation of the generalized Jarzynski equality.
\newblock \emph{Nature Physics} 6:988--992.

\bibitem[B\'{e}rut et~al.(2012)B\'{e}rut, Arakelyan, Petrosyan, Ciliberto,
  Dillenschneider, and Lutz]{Berut2012}
B\'{e}rut, A., A.~Arakelyan, A.~Petrosyan, S.~Ciliberto, R.~Dillenschneider,
  and E.~Lutz, 2012.
\newblock {Experimental verification of Landauer's principle linking
  information and thermodynamics}.
\newblock \emph{Nature} 483:187--189.

\bibitem[Mandal and Jarzynski(2012)]{Mandal2012}
Mandal, D., and C.~Jarzynski, 2012.
\newblock {Work and information processing in a solvable model of Maxwellfs
  demon}.
\newblock \emph{Proc. Nat. Acad. Sci.} 109.

\bibitem[Horowitz et~al.(2013)Horowitz, Sagawa, and Parrondo]{Horowitz2012}
Horowitz, J.~M., T.~Sagawa, and J.~M.~R. Parrondo, 2013.
\newblock {Imitating chemical motors with optimal information motors}.
\newblock \emph{Phys. Rev. Lett.} 111.

\bibitem[Boyer(1997)]{Boyer1997}
Boyer, P.~D., 1997.
\newblock {The ATP synthase--a splendid molecular machine}.
\newblock \emph{Ann. Rev. Biochem.} 66:717--749.

\bibitem[Yasuda et~al.(1998)Yasuda, Noji, Kinosita, and Yoshida]{yasuda1998}
Yasuda, R., H.~Noji, K.~Kinosita, and M.~Yoshida, 1998.
\newblock {F$_1$-ATPase is a highly efficient molecular motor that rotates with
  discrete 120$^\circ$ steps}.
\newblock \emph{Cell} 93:1117--1124.

\bibitem[Wang and Oster(1998)]{Wang1998}
Wang, H., and G.~Oster, 1998.
\newblock {Energy transduction in the F$_1$ motor of ATP synthase.}
\newblock \emph{Nature} 396:279--282.

\bibitem[Itoh et~al.(2004)Itoh, Takahashi, Adachi, Noji, Yasuda, Yoshida, and
  Kinosita]{itoh2004}
Itoh, H., A.~Takahashi, K.~Adachi, H.~Noji, R.~Yasuda, M.~Yoshida, and
  K.~Kinosita, 2004.
\newblock {Mechanically driven ATP synthesis by F$_1$-ATPase}.
\newblock \emph{Nature} 427:465--468.

\bibitem[Toyabe et~al.(2010{\natexlab{b}})Toyabe, Okamoto, Watanabe-Nakayama,
  Taketani, Kudo, and Muneyuki]{toyabe2010_F1}
Toyabe, S., T.~Okamoto, T.~Watanabe-Nakayama, H.~Taketani, S.~Kudo, and
  E.~Muneyuki, 2010.
\newblock {Nonequilibrium energetics of a single F$_1$-ATPase molecule}.
\newblock \emph{Phys. Rev. Lett.} 104:198103--198106.

\bibitem[Toyabe and Muneyuki(2012)]{toyabe2013}
Toyabe, S., and E.~Muneyuki, 2012.
\newblock {Nanosized free-energy transducer F$_1$-ATPase achieves 100\%
  efficiency at finite time operation.} Preprint at
  http://arxiv.org/abs/1210.4017.

\bibitem[Evans et~al.(1993)Evans, Cohen, and Morriss]{evans1993}
Evans, D.~J., E.~G.~D. Cohen, and G.~P. Morriss, 1993.
\newblock Probability of second law violations in shearing steady states.
\newblock \emph{Phys. Rev. Lett.} 71:2401--2404.

\bibitem[Jarzynski(1997)]{jarzynski1997}
Jarzynski, C., 1997.
\newblock Nonequilibrium equality for free fnergy differences.
\newblock \emph{Phys. Rev. Lett.} 78:2690--2693.

\bibitem[Kurchan(1998)]{kurchan1998}
Kurchan, J., 1998.
\newblock {Fluctuation theorem for stochastic dynamics}.
\newblock \emph{J. Phys. A} 31:3719--3729.

\bibitem[Watanabe et~al.(2012)Watanabe, Okuno, Sakakihara, Shimabukuro, Iino,
  Yoshida, and Noji]{watanabe2012}
Watanabe, R., D.~Okuno, S.~Sakakihara, K.~Shimabukuro, R.~Iino, M.~Yoshida, and
  H.~Noji, 2012.
\newblock {Mechanical modulation of catalytic power on F$_1$-ATPase}.
\newblock \emph{Nat. Chem. Biol.} 8:86--92.

\bibitem[Adachi et~al.(2012)Adachi, Oiwa, Yoshida, Nishizaka, and
  Kinosita]{adachi2012}
Adachi, K., K.~Oiwa, M.~Yoshida, T.~Nishizaka, and K.~Kinosita, 2012.
\newblock {Controlled rotation of the F$_1$-ATPase reveals differential and
  continuous binding changes for ATP synthesis.}
\newblock \emph{Nat. Commun.} 3:1022--1033.

\bibitem[Zimmermann and Seifert(2012)]{zimmermann2012}
Zimmermann, E., and U.~Seifert, 2012.
\newblock {Efficiencies of a molecular motor: a generic hybrid model applied to
  the F$_1$-ATPase}.
\newblock \emph{New J. Physics} 14:103023--103042.

\bibitem[Masaike et~al.(2008)Masaike, Koyama-Horibe, Oiwa, Yoshida, and
  Nishizaka]{masaike2008}
Masaike, T., F.~Koyama-Horibe, K.~Oiwa, M.~Yoshida, and T.~Nishizaka, 2008.
\newblock {Cooperative three-step motions in catalytic subunits of F$_1$-ATPase
  correlate with 80 degrees and 40 degrees substep rotations.}
\newblock \emph{Nat. Struct. Mol. Biol.} 15:1326--1333.

\bibitem[Yasuda et~al.(2001)Yasuda, Noji, Yoshida, Kinosita, and
  Itoh]{yasuda2001}
Yasuda, R., H.~Noji, M.~Yoshida, K.~Kinosita, and H.~Itoh, 2001.
\newblock {Resolution of distinct rotational substeps by submillisecond kinetic
  analysis of F$_1$-ATPase.}
\newblock \emph{Nature} 410:898--904.

\bibitem[Toyabe et~al.(2012)Toyabe, Ueno, and Muneyuki]{toyabe2012}
Toyabe, S., H.~Ueno, and E.~Muneyuki, 2012.
\newblock Recovery of state-specific potential of molecular motor from
  single-molecule trajectory.
\newblock \emph{Euro. Phys. Lett.} 97:40004--40009.

\bibitem[Toyabe et~al.(2011)Toyabe, Watanabe-Nakayama, Okamoto, Kudo, and
  Muneyuki]{toyabe2011stall}
Toyabe, S., T.~Watanabe-Nakayama, T.~Okamoto, S.~Kudo, and E.~Muneyuki, 2011.
\newblock {Thermodynamic efficiency and mechanochemical coupling of
  F$_1$-ATPase}.
\newblock \emph{Proc. Natl. Acad. Sci} 108:17951--17956.

\bibitem[Okuno et~al.(2010)Okuno, Iino, and Noji]{okuno2010}
Okuno, D., R.~Iino, and H.~Noji, 2010.
\newblock {Stiffness of $\gamma$ subunit of F$_1$-ATPase.}
\newblock \emph{Eur. Bio. Phys. J} 39:1589--1596.

\bibitem[Harada and Sasa(2005)]{haradasasa2005}
Harada, T., and S.-I. Sasa, 2005.
\newblock Equality connecting energy dissipation with a violation of
  fluctuation-response relation.
\newblock \emph{Phys. Rev. Lett.} 95:130602--130605.

\bibitem[Iko et~al.(2009)Iko, Tabata, Sakakihara, Nakashima, and Noji]{Iko2009}
Iko, Y., K.~V. Tabata, S.~Sakakihara, T.~Nakashima, and H.~Noji, 2009.
\newblock {Acceleration of the ATP-binding rate of F1-ATPase by forcible
  forward rotation}.
\newblock \emph{FEBS lett.} 583:3187--3191.

\bibitem[Iwaki et~al.(2009)Iwaki, Iwane, Shimokawa, Cooke, and
  Yanagida]{Iwaki2009}
Iwaki, M., A.~H. Iwane, T.~Shimokawa, R.~Cooke, and T.~Yanagida, 2009.
\newblock {Brownian search-and-catch mechanism for myosin-VI steps.}
\newblock \emph{Nat. Chem. Biol.} 5:403--405.

\bibitem[DeWitt et~al.(2012)DeWitt, Chang, Combs, and Yildiz]{DeWitt2012}
DeWitt, M.~A., A.~Y. Chang, P.~A. Combs, and A.~Yildiz, 2012.
\newblock {Cytoplasmic dynein moves through uncoordinated stepping of the AAA+
  ring domains.}
\newblock \emph{Science} 335:221--225.

\bibitem[Marcus(1993)]{Marcus1993}
Marcus, R.~A., 1993.
\newblock Electron transfer reactions in chemistry. Theory and experiment.
\newblock \emph{Rev. Mod. Phys.} 65:599--610.

\end{thebibliography}
\bibliographystyle{biophysj}








\newpage

\begin{center}
\title{\LARGE{Supporting Material}} 
\end{center}

\renewcommand{\figurename}{FIGURE}
\makeatletter 
\renewcommand{\thefigure}{S\arabic{figure}} 
\renewcommand{\theequation}{S\arabic{equation}} 
 \renewcommand{\refname}{SUPPORTING REFERENCES}
\setcounter{equation}{0}

\setcounter{figure}{0}

\section*{A. Effective potential in the fast chemical reaction limit}

Here we show that by assuming fast ATP hydrolysis and Pi release, the dynamics apart from the slow $120^\circ$ steps are governed by the overdamped Brownian motion inside the effective potentials $\Pot{i}{x}$.
Through the same method, we may prove that in the high ATP concentration limit, one obtains the tilted periodic potential picture (Fig.~2{\it C} in the main text) for the full dynamics.

\begin{figure}[!bp]
 \begin{center}
  \includegraphics[width=150mm]{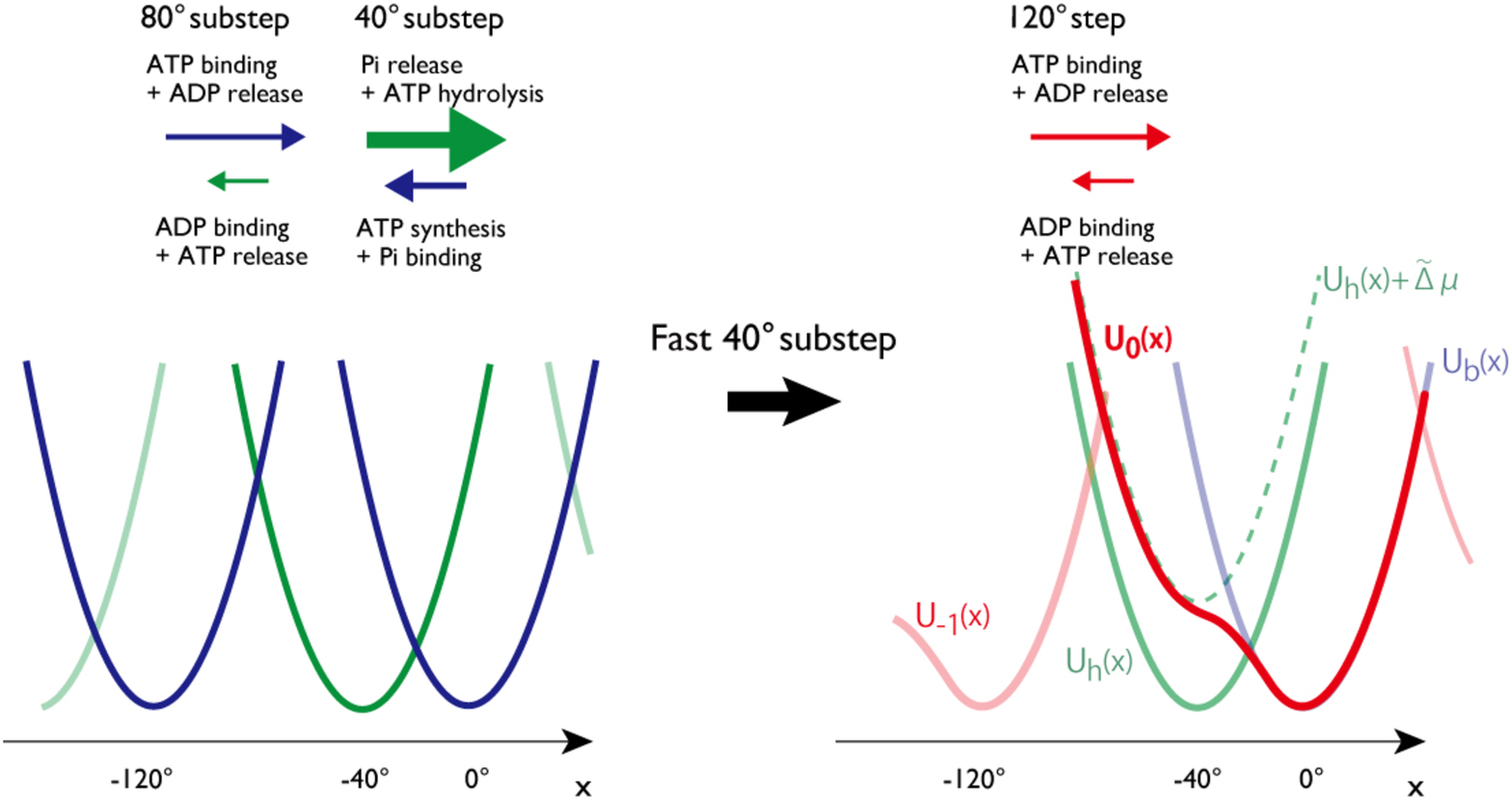}
 \end{center}
  \caption{\label{S1} Left: Mechanical potentials and chemical reactions correponding to the substeps. Right: In the limit of the fast 40$^\circ$ substep, the two potentials, $U_{\rm h}(x)$ and $U_{\rm b}(x)$, corresponding to the ATP hydrolysis dwell and the ATP binding dwell, respectively, will be merged into one effective potential, $U_0(x)$. When $U_{\rm h}(x)$ and $U_{\rm b}(x)$ are assumed to be harmonic with the same spring constant (as observed in \cite{watanabe2012}), $U_0(x)$  is given by Eq.~(\ref{newpot}). Numerical results presented in the main text were obtained using this $U_0(x)$. }
\end{figure}

We consider that there are two potentials (Fig.~\ref{S1}), $U_{\rm h}(x)$ and $U_{\rm b}(x)$, corresponding to the ATP hydrolysis dwell (centered at $x=-l = -40^\circ$) and the ATP binding dwell ($x=0$).
To neglect the slow switching (80$^\circ$ step), we assume that the probe is contained in either of the potential for the time scale of interest.
We assume that the potential energy is large compared to the thermal energy, $U_{\rm h}(0),U_{\rm b}(-l) \gg k_{\rm B} T$, which is the case observed in experiment.
Let $\Prob{\rm h}{x}$ and $\Prob{\rm b}{x}$ be the probability density functions of finding $x$ inside each potentials.
The Fokker-Planck equations read
\begin{eqnarray}
\frac{\partial}{\partial \tilde{t}} \Probt{{i}}{\tilde{x}} &=&  \frac{k_{\rm B} T}{\Gamma X l_v^2} \frac{\partial}{\partial \tilde{x}} \left[\frac{d \WPot{{ i}}{\tilde{x}}}{d \tilde{x}} \Probt{{i}}{\tilde{x}}  +  \frac{\partial}{\partial \tilde{x}} \Probt{{i}}{\tilde{x}} \right] +   \Swt{j}{\tilde{x}} \Probt{j}{\tilde{x}}  - \Swt{\rm i}{\tilde{x}} \Probt{i}{\tilde{x}} , \label{FPhb}
\end{eqnarray}
where $i$ and $j\neq i$ are h or b.
We have normalized the equation using $\tilde{t} = X t $, $\tilde{x} = x/l_v$, and $\WPot{i}{x}=\Pot{{i}}{x}/k_{\rm B} T$.
Here, $X$ is the typical (slowest) rate of the ATP hydrolysis or the Pi releasing reaction.
As the typical length scale $l_v$, we shall adopt the length scale of $E(x):= [\Pot{\rm h}{x}- \Pot{\rm b}{x} + \widetilde{\Delta} \mu ] / k_{\rm B} T$, which is much smaller than the length scale of $\Pot{\rm h}{x}$ in the large potential energy setup.
The switching rates from $h$ to $b$ and $b$ to $h$ have been defined as $X \Swt{{h}}{\tilde{x}}$ and $X \Swt{{b}}{\tilde{x}}$, respectively, which satisfy the local detailed balance:
\begin{eqnarray}
\frac{\Swt{\rm h}{x}}{\Swt{\rm b}{x}} = \exp \left\{ \frac{1}{k_{\rm B} T}[ \Pot{\rm h}{x}- \Pot{\rm b}{x} + \widetilde{\Delta} \mu ] \right\}. \label{DB2}
\end{eqnarray}
The free energy difference between the ATP bound state and the ATP hydrolyzed+Pi released state of the F$_1$ is denoted as $\widetilde{\Delta} \mu$.

Assuming fast reaction ($X \to \infty$) corresponds to taking $\epsilon := 1/\tau_v X$ as the small parameter, where $\tau_v:= \Gamma l_v^2/ k_{\rm B} T$.
Let us calculate $\Prob{i}{x}$ in the form, $\Prob{i}{x} = \Probz{i}{x} + \epsilon \Probo{i}{x} + O(\epsilon ^2)$.
We obtain from the 0-th order equations in Eq.~\ref{FPhb}:
\begin{eqnarray}
\Probz{i}{x} = Q^t(x) \Probs{i}{x} \label{Product},
\end{eqnarray}
where $\Probs{\rm h}{x} := \{1+ \exp [E(x)] \}^{-1}$ and $\Probs{\rm b}{x} := \{1+ \exp [-E(x)]\} ^{-1}\  [= 1-\Probs{\rm h}{x}]$.
We adopted the length scale of $E(x)$ as $l_v$ in Eq.~\ref{FPhb} since this is the length scale of $\Probs{\rm h,b}{x}$, which is critical in the perturbation theory.
The solvability condition for the 1st order equations in Eq.~\ref{FPhb} determines the dynamics of $Q^t(x)$:
\begin{eqnarray}
 \frac{\partial}{\partial t}Q^t(x) =  \frac{1}{\Gamma} \frac{\partial}{\partial x} \left[ \sum_{i={\rm h,b}} \Probs{i}{x} \frac{d \Pot{i}{x}}{d x} Q^t(x) + k_{\rm B} T \frac{\partial}{\partial x} Q^t(x)  \right],
\end{eqnarray}
which is equivalent to the one-dimensional overdamped Langevin equation with the effective force
\begin{eqnarray}
-\frac{d \Pot{0}{x}}{ dx }  = -\sum_{i=h,b} \Probs{i}{x} \frac{d \Pot{i}{x}}{d x},
\end{eqnarray}
 where the effective potential is obtained by
\begin{eqnarray}
\Pot{0}{x} := \int ^x_c dx \sum_{i=h,b} \Probs{i}{x} \frac{d \Pot{i}{x}} {d x},
\end{eqnarray}
with an arbitrary fixed constant $c$ (Fig.~\ref{S1}.)
Assuming that the two potentials $U_b(x)$ and $U_h(x)$ are harmonic with the same spring constants, $U_b(x) = U_h(x + l) = kx^2/2$, which is consistent with the ATP binding dwell and catalytic dwell observed in experiment~\cite{watanabe2012}, we have an explicit form
\begin{eqnarray}
\Pot{0}{x} = k_{\rm B} T \left\{ \frac{1}{2}k x^2 - \log \left[ e^ {-klx} + e^{ \widetilde{\Delta} \mu / k_{\rm B} T + kl^2/2} \right] \right\}. \label{newpot}
\end{eqnarray}
We used Eq.~\ref{newpot} to fit the potential estimated from the probe trajectory~\cite{toyabe2012} by the parameters $k$ and $\widetilde{\Delta} \mu$, and obtained $k=0.0061$ deg$^{-2}$ and $\widetilde{\Delta} \mu = 5.2 k_{\rm B} T$.

The high ATP concentration case of the full dynamics [which consists of potentials $U_n(x)$ and switching rates $\Swtpm{n}{x}$] could be treated in a similar manner if we adopt as $l_v$ the length scale of $E_n(x):=[\Pot{n}{x}- \Pot{n+1}{x} + \Delta {\mu} ] / k_{\rm B} T$, and consider the limit $W\tau_v \gg 1$.
The dynamics in this limit is described by
 \begin{eqnarray}
 \Gamma \dot{x} = F(x) + \sqrt{2\Gamma k_{\rm B} T} \xi_t. \label{1dLangevin}
 \end{eqnarray}
The effective force $F(x)$ is given by
\begin{eqnarray}
F(x) =  - \sum_{n=-\infty} ^{\infty} \Probs{n}{x} \frac{d \Pot{n}{x}}{d x}, \label{eff_force}
\end{eqnarray}
with $\Probs{n}{x}$ defined similarly to the previous case as
\begin{eqnarray}
\Probs{n}{x} := \frac{\exp \left\{ -[\Pot{n}{x}-n\Delta \mu ]/k_{\rm B} T \right\} }{\sum_{m=-\infty}^\infty \exp \left\{ -[\Pot{m}{x}-m\Delta \mu ]/k_{\rm B} T \right\} }.
\end{eqnarray}
The force in Eq.~\ref{eff_force} corresponds to a tilted periodic potential, where the energy difference per 120$^\circ$ step is $\Delta \mu$ (Fig.~2{\it B}).
Since this energy difference is dissipated through the rotational motion of the probe,
\begin{eqnarray}
\Qext = -\int_0 ^{120^\circ} F(x) dx = \Delta \mu.
\end{eqnarray}
The maximum velocity $v$, which is the steady-state velocity of the model described by \ref{1dLangevin}, may be obtained analytically using $F(x)$.

\section*{B. Harmonic potential model}

We consider in this section the simplified harmonic potential case, $\Pot{n}{x}=K(x-nL)^2/2$, with $L=120^\circ$.
In Fig.~\ref{vQharmonic}, we show the numerical results of $\Qext$ in this model.
Under the condition that the diffusion coefficient $D=k_{\rm B} T / \Gamma$ as $D/L^2 =3.3 {\rm sec}^{-1}$ \cite{toyabe2012}, and the chemical potential as $\Delta \mu/ k_{\rm B} T=19$ \cite{toyabe2010_F1, toyabe2011stall}, the value of $K$ was determined as $KL^2/k_{\rm B} T = 50$ by setting the maximum average velocity to fit with that obtained in experiment.
The characteristic feature of $q$ dependence is similar to the case of Fig.~2{\it C}, where the potential estimated through experiment was used in the calculation.
Note that in this model, the angular position dependence of the forward switching rate has a simple form,
\begin{eqnarray}
R^+_n(x) \propto \exp \left[ qKLx / k_{\rm B} T \right],   \label{R_harmonic}
\end{eqnarray}
which allows us to directly compare the value of $q$ with the experimentally observed rates of ATP binding \cite{watanabe2012,Iko2009,adachi2012}. Using the above parameters, we obtain $q =0.07\sim0.1$, $0.11$, and $0.12$ for \cite{Iko2009}, \cite{watanabe2012}, and \cite{adachi2012}, respectively, which is consistent with our observation that $q$ should be close to zero in order to explain the internal dissipation-free and asymmetric velocity features of the F1 motor.

\begin{figure}[!bp]
 \begin{center}
  \includegraphics[width=75mm]{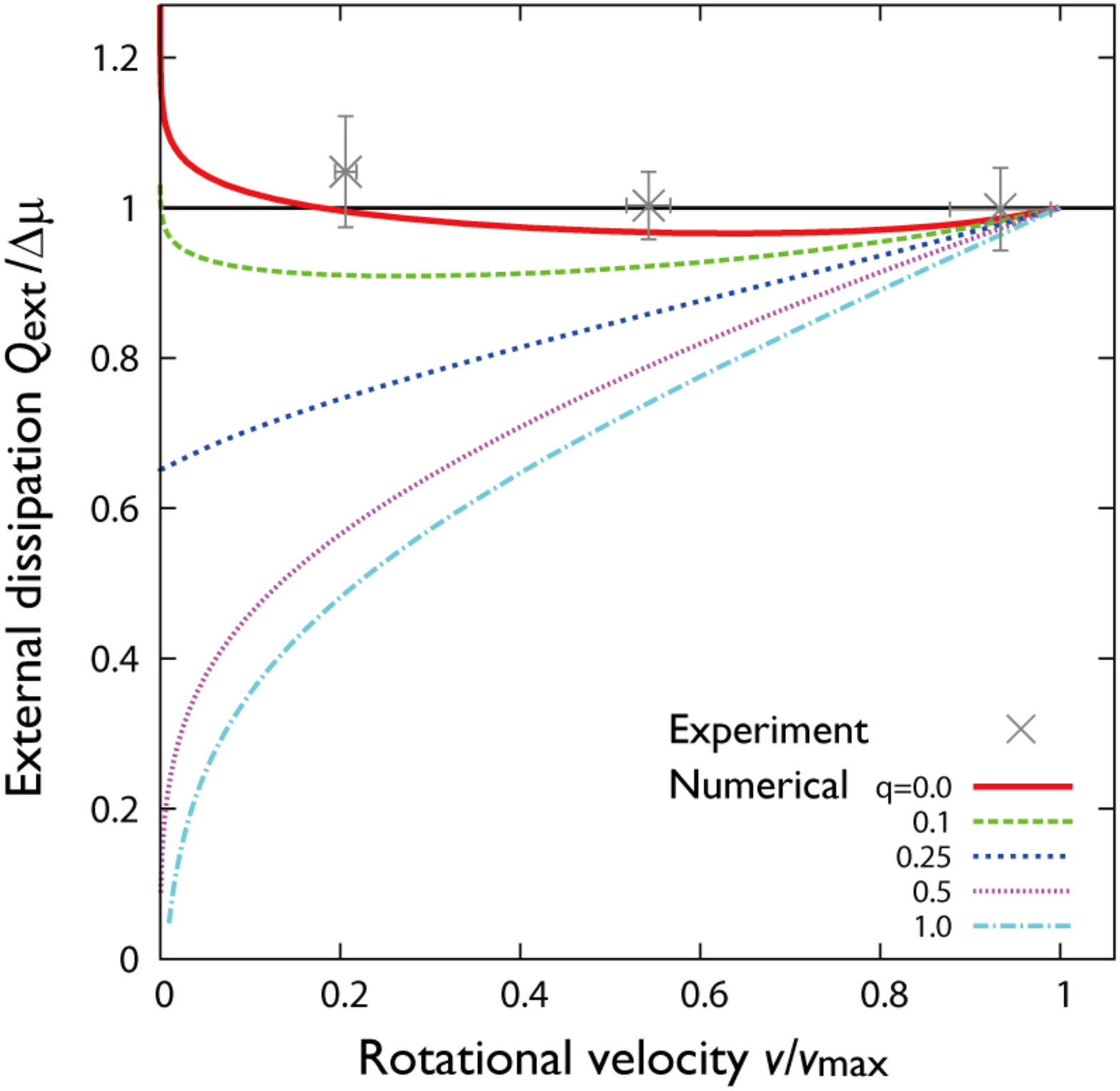}
 \end{center}
  \caption{\label{vQharmonic} Rotational velocity $v$ versus the external heat dissipation per step $\Qext$ in the harmonic potential model. Parameters are given in the text. The experimental results were obtained from~\cite{toyabe2010_F1} (error bar: standard error of mean).}
\end{figure}

Let us first consider the large $W$ limit (high ATP concentration). The length scale of potentials $\Pot{n}{x}=K(x-nL)^2/2$ and that of  $\Pot{n}{x}-\Pot{n \pm 1}{x}= \mp K Lx$ are $\sqrt{k_{\rm B} T/ K}$ and $k_{\rm B} T/ (KL)$, respectively. 
Since the potential energy is sufficiently large $KL^2/k_{\rm B} T \gg 1$, the smallest length scale in this model is $l_v = k_{\rm B} T/ KL$. This length defines the time scale $\tau_v= \Gamma k_{\rm B} T /(KL)^2$, which determines the typical $W$ (ATP concentration) that allows the effective force description of the model, and consequently the velocity saturation.
Let us also define $\tau_{\rm p}:=  \Gamma / K$ $(\gg \tau_v)$, which corresponds to the time scale of equilibration inside a single potential.

Significance of the time scale $\tau_v$ is numerically verified through seeing how the velocity dependence of $W$ in the model changes according to the spring constant $K$.
In Fig.~\ref{VelSat}, we show the results for the case where $K$ and $\Delta \mu$ are parameterized by $d$ $(= -1,0,1,2,3,4,5)$ as
\begin{eqnarray}
KL^2/k_{\rm B} T &=& 50 \times 2^d \label{scaleK} \\ 
\Delta \mu / k_{\rm B} T &=& 19 \times 2^d.  \label{scaleMu}
\end{eqnarray}
Clearly, the value of $W$ at which the velocity saturates is scaled by $\tau_v$ ($\propto K^{-2}$) and not by $\tau_{\rm p}$ ($\propto K^{-1}$), when $d$ is sufficiently large.

\begin{figure}[!tbp]
 \begin{center}
  \includegraphics[width=80mm]{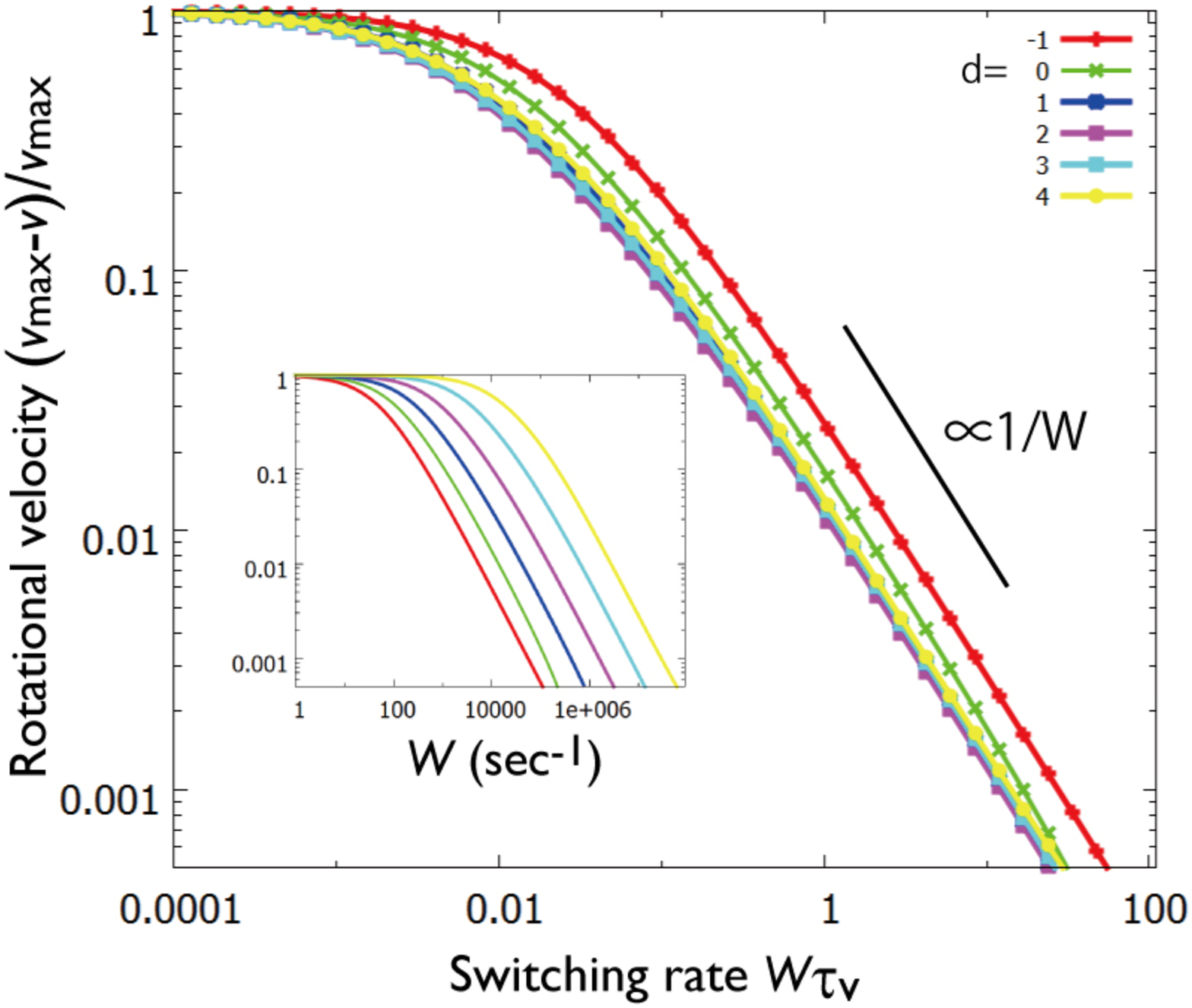}
 \end{center}
  \caption{\label{VelSat} $W$ dependence of velocity for the harmonic potential model with $q=0$ (numerical). Different colors correspond to different $d$'s, which changes the set of spring constant $K$ and hydrolysis free energy $\Delta \mu$ in the model according to Eqs.~\ref{scaleK} and \ref{scaleMu}. Results from models with different $d$ are plotted by scaling $W$ by $\tau_v ^{-1}$. Inset shows same data without scaling $W$.}
\end{figure}

To understand the limit $W\to 0$ of the model, we focus on the switching dynamics between potentials $U_0(x)$ and $U_1(x) -\Delta \mu$, since the dynamics between two neighboring potentials
are equivalent in steady-state.
Our aim is to estimate the probability density of the position where the switching from $U_0(x)$ to $U_1(x) -\Delta \mu$ takes place:
\begin{eqnarray}
\Trans{0}{x} := \Lambda_0(x)/W = \Probss{0}{x} f_0^+(x) - \Probss{1}{x} f_1^-(x).
 \label{trans}
\end{eqnarray}
$\Probss{0}{x}$ and $\Probss{1}{x}$ are the steady-state densities of $x$ under the condition that $n$ is 0 and 1, respectively.
The first term in the right-hand side of Eq.~\ref{trans} corresponds to the probability density of the switching at $x$, whereas the second part is that of the switch back (1$\to$0).
When $\Trans{0}{x}$ is obtained, the internal heat dissipation may be calculated as
\begin{eqnarray}
\Qint = \frac{1}{Z} \int dx \lambda_0(x) \left[ \Pot{0}{x}-\Pot{1}{x} + \Delta \mu \right],
\end{eqnarray}
where $Z = \int dx \lambda_0(x)$ is the normalization factor.

For $W \ll \tau_{\rm p}^{-1}$, the steady-state probability density of $x$ is close to the equilibrium density inside each potential
\begin{eqnarray}
\Probss{n}{x} \simeq \Probeq{n}{x}\propto \exp [-U_n(x)/k_{\rm B} T].\label{equilibrium}
\end{eqnarray}
Although this assumption is valid in estimating the first term in the right-hand side of Eq.~\ref{trans}, it fails to capture the feature of the second term at $W > 0$, since the small but finite switching makes $\Probss{1}{x}$ deviate from $\Probeq{1}{x}$ at around the peak point of $\Probss{0}{x} f^+_0 (x)$, where $f^-_1 (x)$ may take a large value.

We focus on the model with $q<x_c/L$, where $x_c:= (KL^2/2 - \Delta \mu)/ k_{\rm B}T KL \simeq 14^{\circ}$ is the intersection point between the two potentials, $U_0(x_c)-U_1(x_c)+\Delta \mu = 0$.
In this region of $q$, $\Probeq{0}{x} f^+_0 (x)$ has a peak at $x<x_c$.
In order to phenemenologically take into account the effect of switch back, we consider the conditional probability that after the switching occurs at $x$, the potential stays as $U_1(x)-\Delta \mu$ and is not switched back to $U_0(x)$:
\begin{eqnarray}
D_0(x)&:=& \frac{\exp[-\tau_v/\tau_{\rm leq}(x)] + \exp[E_0(x)]}{1+ \exp[E_0(x)]}. \label{switch}
\end{eqnarray}
We have introduced the local equilibrium time scale
\begin{eqnarray}
\tau_{\rm leq}(x) := \frac{1}{\Swtp{0}{x}+\Swtm{1}{x} } = \frac{1}{W [\fSwtp{0}{x}+\fSwtm{1}{x}]},
\end{eqnarray}
which is the typical time required for equilibration between $U_0(x)$ and $U_1(x)-\Delta \mu$ at a fixed position $x$.
Using $D_0(x)$, we assume that the switching position probability density is given by
\begin{eqnarray}
\tilde{\lambda}_0(x) := \Probeq{0}{x} f^+_0 (x) D_0(x).  \label{PTrans}
\end{eqnarray}
This is justified since the main contribution from the $\Probss{1}{x} f^-_1 (x)$ term in Eq.~\ref{trans} is the switch back which occurs right after the switch 0$\to$1, and the probability that the probe spontaneously climbs the potential $U_1(x)$ in the backward direction for the switch back to occur is negligibly small.

\begin{figure}[!t]
 \begin{center}
  \includegraphics[width=80mm]{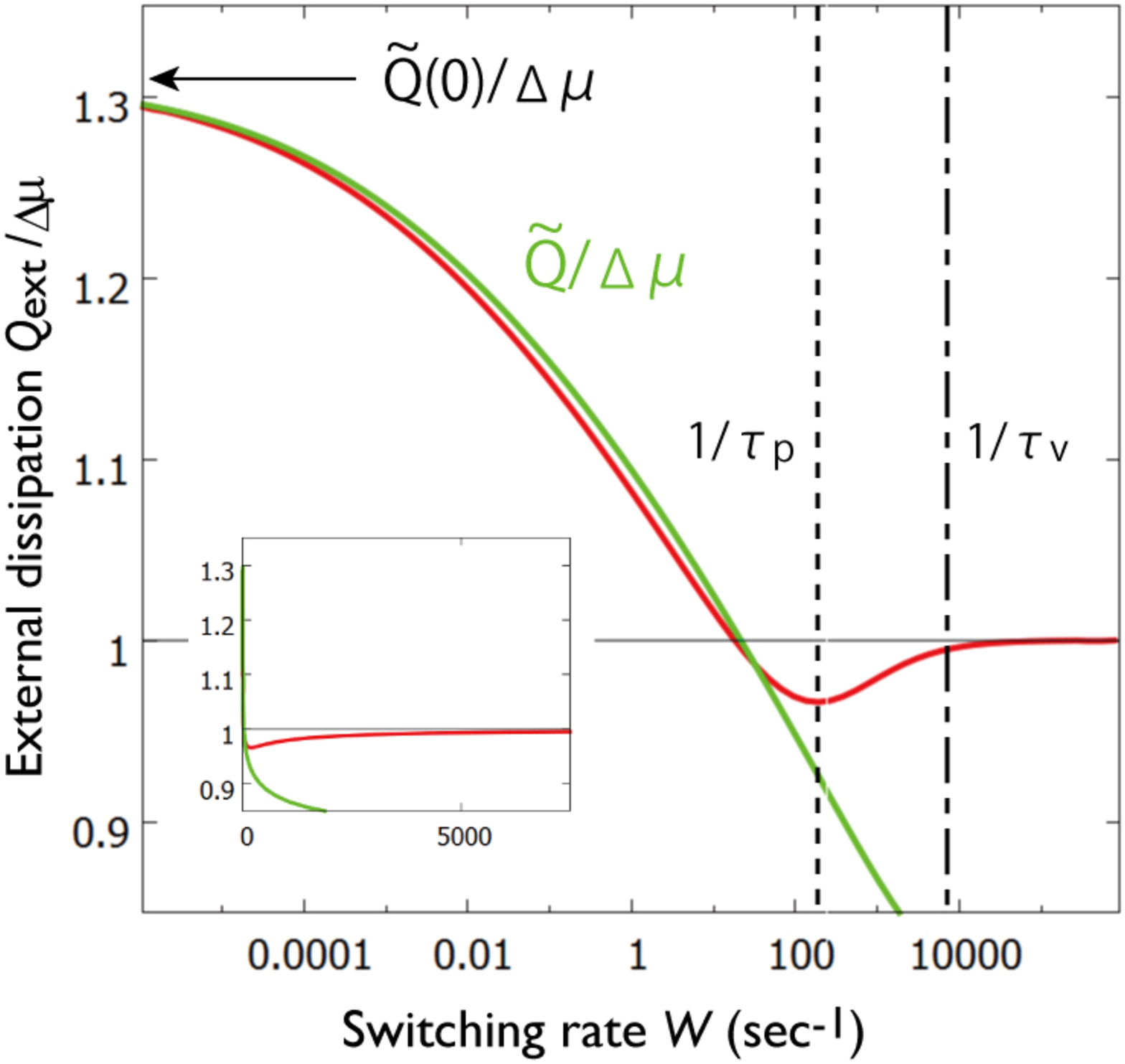}
 \end{center}
  \caption{\label{WQ} Numerically obtained $\Qext$ (red) and the theoretical $\tilde{Q}$ (green) obtained from Eq.~\ref{pheno_qext} in the $q=0$ model. Starting from $\tilde{Q}(0)$ at the limit $W\to 0$, $\Qext$ drops sharply in a manner $\propto -\log W$ at low but finite $W$. $\Qext$ stops dropping at $W\sim \tau_p^{-1}$, and converges to $\Delta \mu$ at $W > \tau_v^{-1}$. Inset shows same data with linear-scale $W$.}
\end{figure}

As shown in Fig.~\ref{WQ}, the external heat dissipation theoretically obtained as
\begin{eqnarray}
\widetilde{Q}(W) &:=& \Delta \mu - \frac{1}{Z} \int dx \tilde{\lambda}_0(x) \left[ \Pot{0}{x}-\Pot{1}{x} + \Delta \mu \right] \\
 &=& \frac{1}{2} K L^2 -  \frac{KL}{Z} \int dx \tilde{\lambda}_0(x) x, \label{pheno_qext}
\end{eqnarray}
captures the feature of $\Qext$ at small $W$.
Note that in the limit $W \to 0$, we find
\begin{eqnarray}
\Qext = \widetilde{Q}(0) = (1/2-q)KL^2, \label{qext_zero}
\end{eqnarray}
since in this limit the switching position probability density becomes $\Probeq{0}{x} f^+_0 (x) \propto \exp [ - K(x-qL ) ^2/2k_{\rm B} T ]$, a Gaussian distribution with peak at $x=qL$.
For finite $W$, the value of $\Qint$ deviates drastically from $\widetilde{Q}(0)$
in a manner $\propto - \log W$, which is observed as a sharp drop when $W$ or $v$ is linear scaled (Fig.~\ref{WQ} inset, Fig.~\ref{vQharmonic}).
Physically, this corresponds to the fact that very little ADP concentration is sufficient to prevent switching to occur at energetically unfavorable positions [$\Pot{1}{x} - \Pot{0}{x} - \Delta \mu \gg k_{\rm B} T$].

\begin{figure}[!t]
 \begin{center}
  \includegraphics[width=80mm]{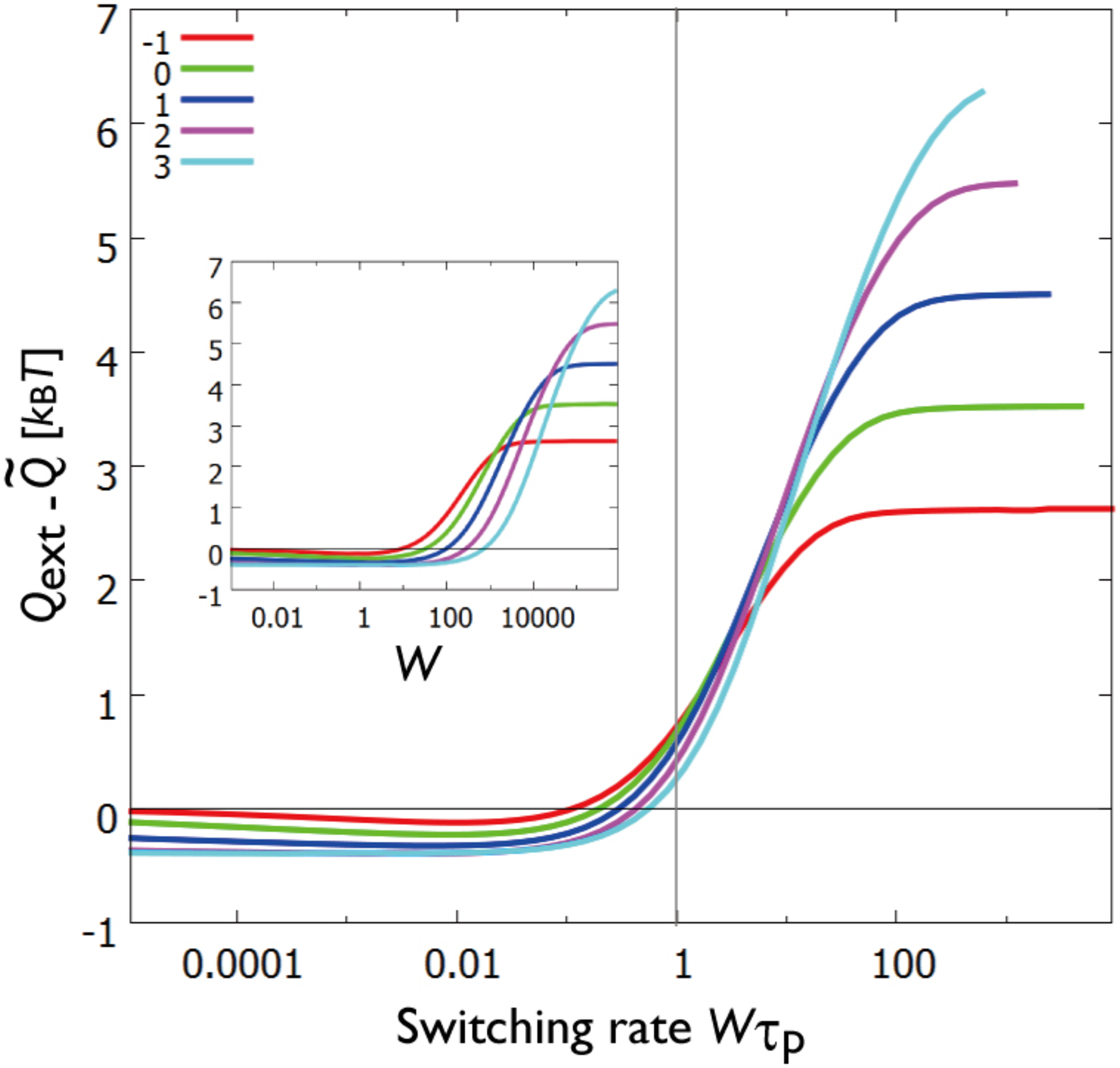}
 \end{center}
  \caption{\label{Tp} Deviation of $\Qext$ from the theoretically obtained $\tilde{Q}(W)$ from Eq.~\ref{pheno_qext}. Different colors correspond to $d=-1,0,1,2,3,4,5$ in the model parameterized by Eqs.~\ref{scaleK}) and \ref{scaleMu}. Inset shows same data without scaling.}
\end{figure}

$\lambda_0(x) \simeq \tilde{\lambda}_0(x)$ is valid when $W \ll \tau_{\rm p}^{-1}$, and should fail when $W > \tau_{\rm p}^{-1}$ since Eq.~\ref{equilibrium} used to evaluate the first term of Eq.~\ref{trans} is violated in this region.
The $\Qext$ therefore deviates from the sharp theoretical curve at around $W \sim \tau_{\rm p}^{-1}$ (Fig.~\ref{Tp}). 
As shown in Fig.~\ref{WQ}, the value of $\tilde{Q}(W)$ is sufficiently close to $\Delta \mu$ when $W \sim \tau_{\rm p}^{-1}$, which could be understood as follows.
Assuming $D(x) \simeq \exp[-\tau_v/\tau_{\rm leq}(x)]$, the peak position $x=x_c -\delta$ of $\Lambda_0(x)$ at $W=\tau_{\rm p}^{-1}$ satisfies
\begin{eqnarray}
\frac{K L (x_c - qL)}{k_{\rm B} T} = \frac{K L \delta}{k_{\rm B} T} - q \exp \left[ -\frac{qKL\delta}{k_{\rm B} T} \right] + (1-q) \exp \left[ \frac{(1-q)K L\delta}{k_{\rm B} T} \right]. \label{deltax}
\end{eqnarray}
Using the $\delta$ obtained in Eq.~\ref{deltax}, $\tilde{Q}(W=\tau_{\rm p}^{-1})$ is estimated as $\simeq \Delta \mu + KL \delta$.
At large $A:=KL^2/k_{\rm B} T$ and $B:=\Delta \mu /k_{\rm B} T = O(A)$, the value of $\delta$ satisfying Eq.~\ref{deltax} scales as $A \delta /L \propto \log A$. Therefore, $\tilde{Q}(W=\tau_{\rm p})/\Delta \mu = 1 + O( \log A /A) $, which means that $\tilde{Q}(W=\tau_{\rm p}) \simeq \Delta \mu$ is satisfied with a small error term under $A\gg 1$.

To sum up, in the potential switching model with the switching rates (3) and (4) in the main text and $q<x_c/L$, $\Qext$ becomes sufficiently close to $\Delta \mu$ at $W \sim \tau_{\rm p} ^{-1}$, when the condition $KL^2, \Delta \mu \gg k_{\rm B} T$ is satisfied.
Since $\tau_{\rm p} =  \tau_{v}KL^2/k_{\rm B} T$,  there exists a time scale separation $\tau_{\rm p} \gg \tau_{v}$, hence at $W \sim \tau_{\rm p}^{-1}$ the velocity is still smaller than the maximum velocity, $v<v_{\rm max}$.
This means that if $KL^2/k_{\rm B}T=50$, which is the case where the maximum velocity is close to the real \Fon, the model shows the $\Qext \sim \Delta \mu$ behavior even when [ATP] is as low as 1/50 of the velocity saturating concentration.
Persistent $\Qext \sim \Delta \mu$ for the broad range of $W > \tau_{\rm p} ^{-1}$ allows the low dependence of $\Qext$ on $v$, which explains the internal dissipation-free feature of \Fons observed in experiment. For the case of models with $q>x_c/L$, it is confirmed that there exists a significant difference between $\Qext$ and $\Delta \mu $ for $W = \tau_{\rm p}^{-1}$, even when $d$ is as large as 5 in the parameterization given by Eqs.~\ref{scaleK} and \ref{scaleMu}.
It is left for future studies to theoretically understand the $q>x_c/L$ models (including $q=0.5$ and 1, Fig.~\ref{q051}).

\begin{figure}[h]
 \begin{center}
  \includegraphics[width=80mm]{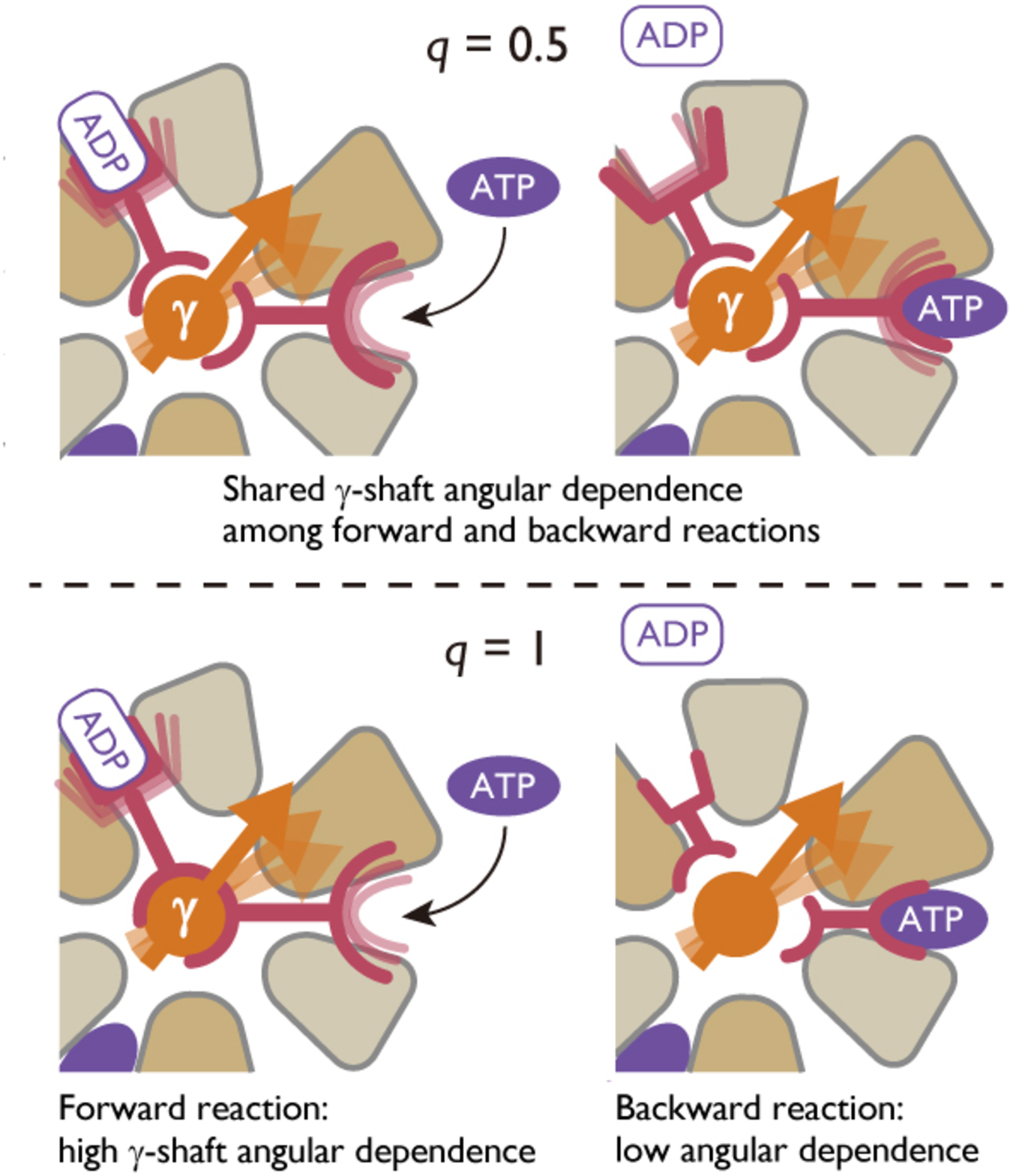}
 \end{center}
  \caption{\label{q051} Schematic of the $q=0.5$ and $q=1$ models. In the $q=0.5$ model, the coordination between the $\gamma$ shaft and the nucleotide binding sites are equally present in the forward and backward reactions. On the other hand, the $\gamma$ shaft and the nucleotide binding sites are only coordinated in the forward step in the $q=1$ model, which is completely opposite to the case of $q=0$ model (Fig.~1{\it B} in the main text). As shown in Fig.~2{\it C} and 3{\it A} in the main text, these models fail to reproduce the internal dissipation-free feature of F$_1$.}
\end{figure}

\newpage
\section*{D. External torque dependence of velocity in various models}

In Fig.~S8, the external torque dependence of the rotational velocity for the $q=0.5$ (left) and 1 (right) models are shown. We used Eq.~\ref{newpot} for $U_0(x)$, with $k=0.0061$ and $\widetilde{\Delta} \mu =5.2 k_B T$. In comparison to the $q=0$ model (Fig.~5A in main text), the $q=0.5$ and 1 models fail to capture the feature of \Fon, where large minus velocity in th  presense of large torque and low nucleotide concentration has been observed.
Notice that in the case of $q=0.5$, the curves are close to anti-symmetric in a wide range of $W$ [represented by $v(0)]$.

We define the intersection switching model by
\begin{eqnarray}
\begin{split}
f_{n}^{+}(x) &=& \exp \left\{ -\frac{(x-x_{c,n})^2}{2\sigma^2} + \frac{q}{k_{\rm B} T}\left[ U_n(x) - U_{n+1} (x)+ \Delta \mu \right ]   \right\}, \\
f_{n+1}^{-} (x) &=& \exp \left\{-\frac{(x-x_{c,n})^2}{2\sigma^2} + \frac{q-1}{k_{\rm B} T}\left[ U_n(x) - U_{n+1} (x) + \Delta \mu \right ] \right\}. 
\end{split}
\label{qmodelgauss}
\end{eqnarray}
Here, $x_{c,n}$ is the intersection point between the two potentials $U_n(x)$ and $U_{n+1}(x)-\Delta \mu$, satisfying $U_{n+1}(x_{c,n})-U_n(x_{c,n}) - \Delta \mu=0$, and $\sigma$ is the typical width of the window of the angle at which the switching is allowed.
If $\sigma$ is sufficiently small, this model would become internal dissipation-free for a wide range of $W$, which seems to explain the experimental data. This is because the switching of the mechanical potential only occurs at angles satisfying $U_{n+1}(x)-U_n(x) - \Delta \mu \sim 0$ in this model. However, if $\sigma$ is too small, the torque dependence of the velocity becomes anti-symmetric with respect to the $F=\Delta \mu /L$ line for all $q$ even at small $W$ (Fig.~S8 left), which is inconsistent with the experimental observations. When $\sigma$ is sufficiently large (Fig.~S8 right), the torque-velocity curve would depend on $q$, which shows that adopting $q \simeq 0$ is critical even in the intersection switching model to reproduce the feature of \Fon.

\begin{figure}[H]
 \begin{center}
  \includegraphics[width=150mm]{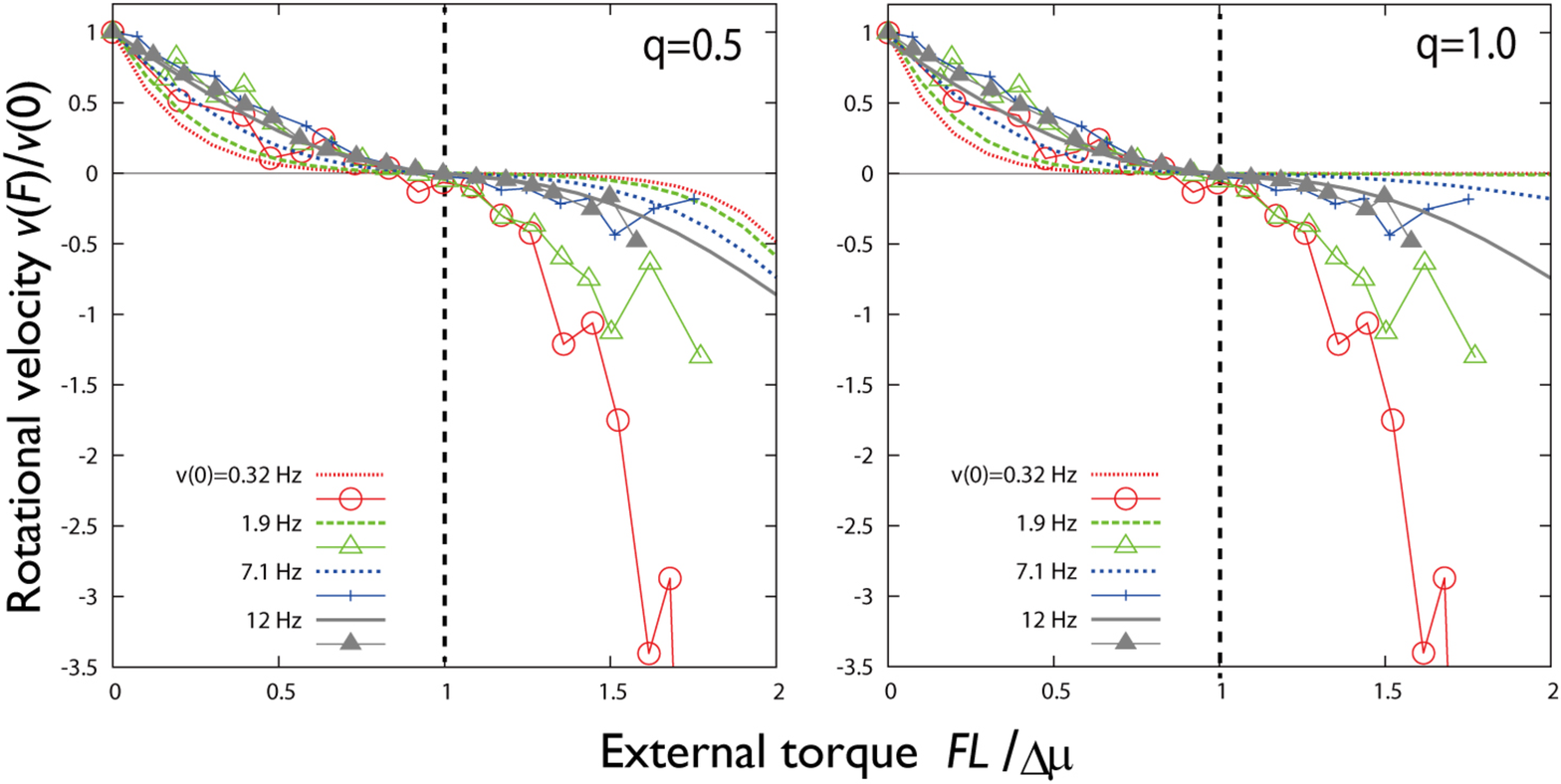}
 \end{center}
  \caption{\label{gauss1} External torque dependence of the rotational velocity in the $q=0.5$ (left) and 1.0 (right) models, plotted with the experimental data~\cite{toyabe2011stall} (kindly provided by S. Toyabe). For each numerical lines, $W$ was chosen and fixed in order to reproduce the values of $v(0)$ of the corresponding experimental data.}
\end{figure}
\begin{figure}[H]
 \begin{center}
  \includegraphics[width=150mm]{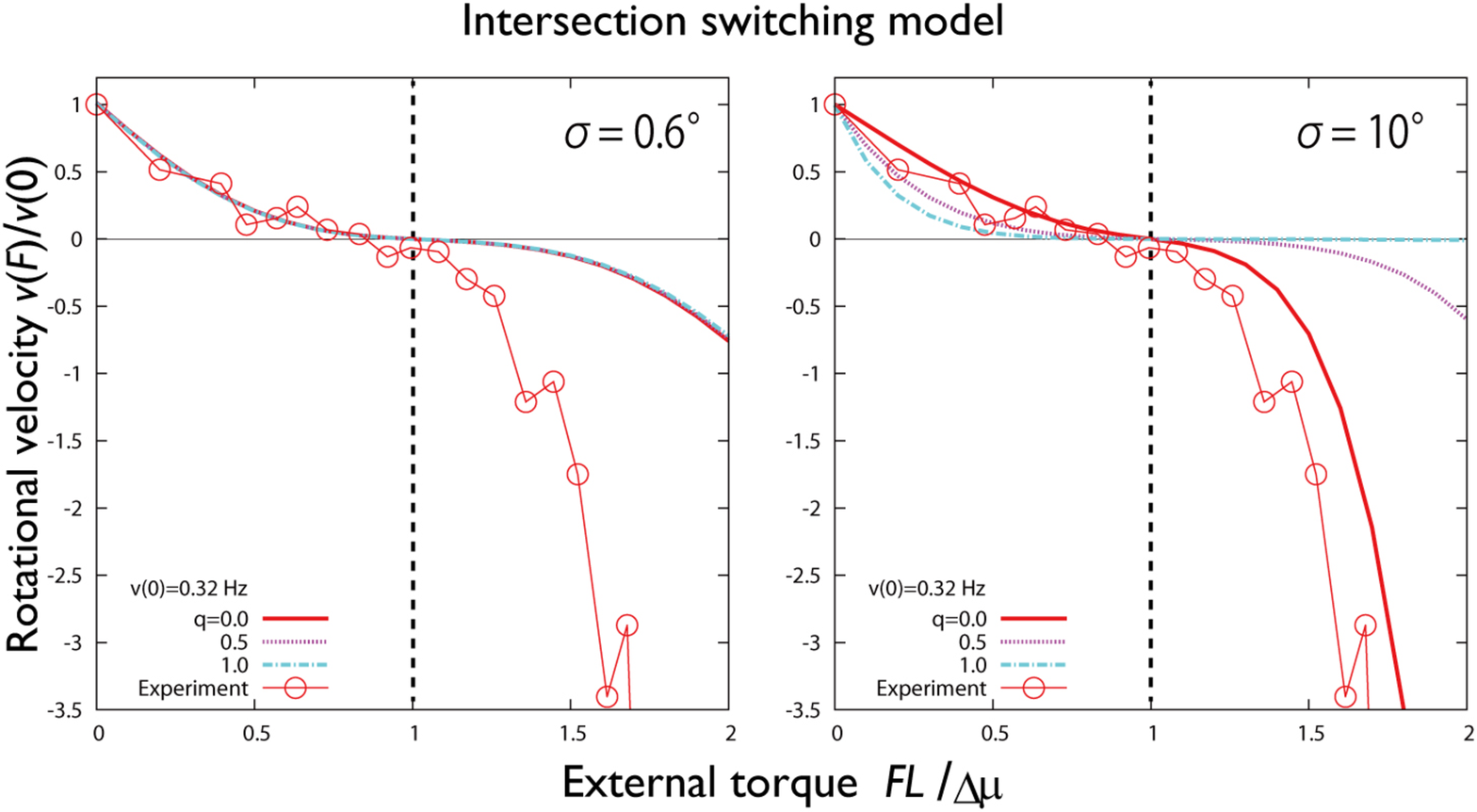}
 \end{center}
  \caption{\label{gauss2} External torque dependence of the rotational velocity for the intersection switching model [see text Eq.~\ref{qmodelgauss}]. When $\sigma$ is small and the switching is only allowed in a narrow range around the potential intersection point (left), the torque-velocity curve becomes anti-symmetric with respect to the $FL=\Delta \mu$ line. When $\sigma$ is set larger (right), the $q$-dependence appears.  For each numerical lines, $W$ was chosen and fixed in order to reproduce the values of $v(0)=0.32$Hz.}
\end{figure}

\begin{figure}[H]
 \begin{center}
  \includegraphics[width=75mm]{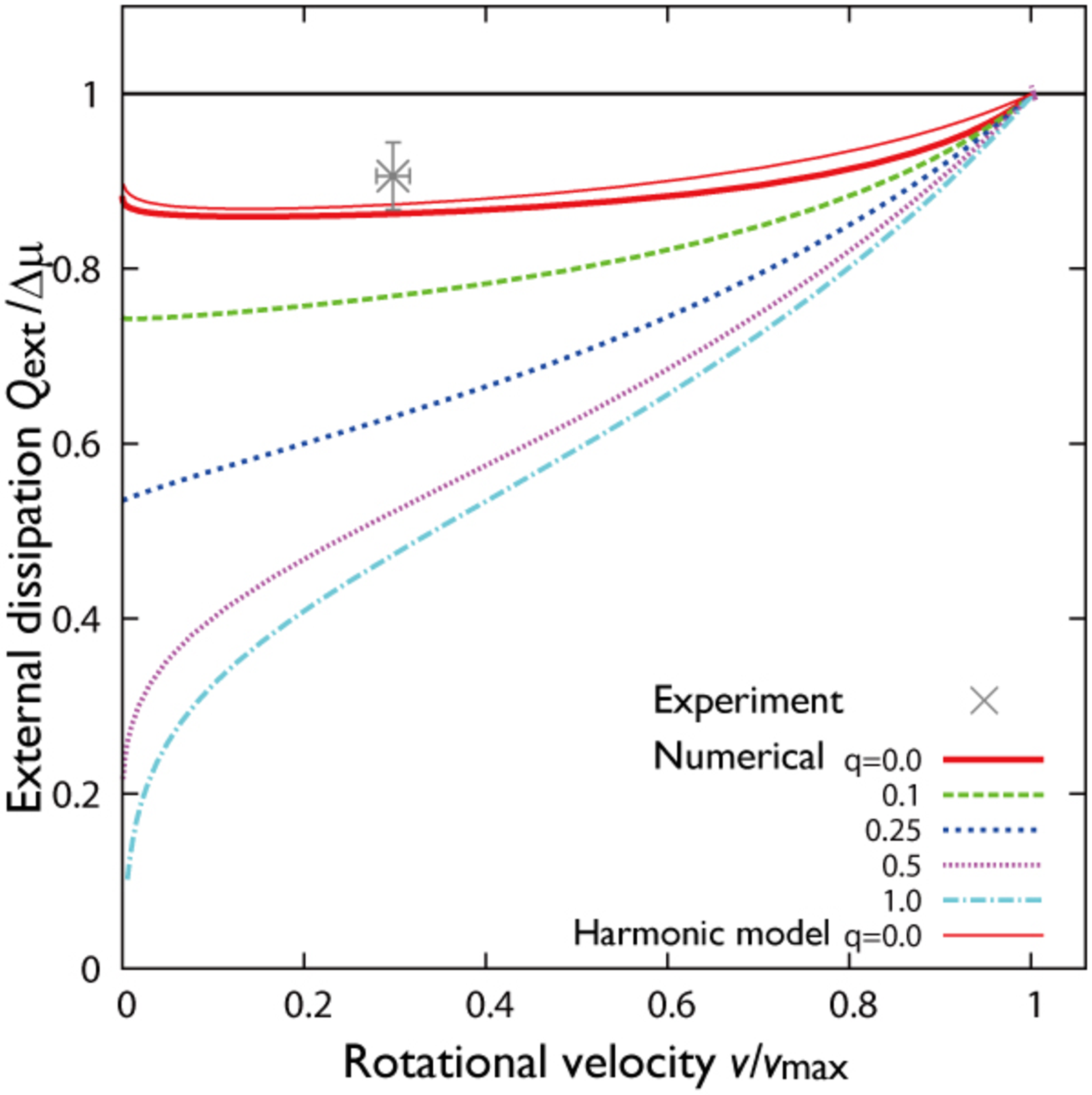}
 \end{center}
  \caption{\label{S9} Rotational velocity $v$ versus the external heat dissipation per step $\Qext$ in the non-harmonic potential model [using Eq.~\ref{newpot} for $U_0(x)$], and the harmonic potential model, in the case of $\Delta \mu = 28 k_{\rm B} T$. Parameters from Supplementary Material A and B were used. The experimental result was obtained from~\cite{toyabe2010_F1} (error bar: standard error of mean). In this large $\Delta \mu$ setup, the intersection point becomes $x_c<0$ in the non-harmonic and harmonic potentials we have introduced. Nevertheless, numerical result for $q=0$ shows consistent value with experiment.}
\end{figure}

\end{document}